\documentclass[a4paper,twoside]{svmult}

\usepackage{natbib}
\bibpunct{(}{)}{;}{a}{,}{,}

\usepackage{makeidx}
\usepackage{multicol}
\makeindex

\usepackage{amsmath}
\usepackage{amsfonts}
\usepackage{mathrsfs} 

\usepackage{graphicx}

\usepackage{color}

\usepackage{endnotes}
\let\footnote=\endnote

\spdefaulttheorem{exemple}{Example}{\bfseries}{\sffamily}
\spnewtheorem*{proof_duplicate}{Proof}{\itshape}{\rmfamily}

\newcommand\path{/home/xian/Books/MCSM2/figures}

\newcommand\Tee{\mathrm{T}}

\begin{document}
\pagenumbering{arabic}
\definecolor{LightGrey}{rgb}{0.86,0.79,0.73}
\renewcommand\labelenumi{\theenumi}
\chapter*{Bayesian computational methods\\
{\sc Christian P. Robert}
{\small{\em Universit\'e Paris-Dauphine, CEREMADE, and CREST, INSEE, Paris}}}
\chaptermark{Bayesian Computations}

\section{Introduction}\label{sec:bayesiancomputation:intro}

If, in the mid 1980's, one had asked 
the average statistician about the difficulties of using Bayesian 
Statistics,\index{Bayesian!Statistics|(}\index{Statistics!Bayesian} 
the most likely answer would have been ``Well, there is this problem
of selecting a prior distribution and then, even if one agrees on the
prior, the whole Bayesian inference is simply impossible to implement
in practice!" The same question asked in the 21th Century does not
produce the same reply, but rather a much less aggressive complaint
about the lack of generic software (besides winBUGS), along with the
renewed worry of subjectively selecting a prior!\index{winBUGS@{\sf winBUGS}} 
The last 20 years have indeed witnessed a tremendous change in the way Bayesian Statistics 
are perceived, both by mathematical statisticians and by applied statisticians
and the impetus behind this change has been a prodigious leap-forward in
the computational abilities. The availability of very powerful approximation
methods has correlatively freed Bayesian modelling, in terms of both model
scope {\em and} prior modelling. This opening has induced many more scientists
from outside the statistics community to opt for a Bayesian perspective as they
can now handle those tools on their own. As discussed below, a most successful
illustration of this gained freedom can be seen in Bayesian model choice,
which was only emerging at the beginning of the MCMC era, for lack
of appropriate computational tools.\index{algorithm!MCMC}

In this chapter, we will first present the most standard computational
challenges met in Bayesian Statistics (Section
\ref{sec:bayesiancomputation:chall}), and then relate these problems
with computational solutions. Of course, this chapter is only a terse 
introduction to the problems and solutions related to Bayesian computations. 
For more complete references, see \cite{robert:casella:2004}, \cite{marin:robert:2007}, \cite{robert:casella:2004} and 
\cite{Liu:2001}, among others. We also restrain from providing an
introduction to Bayesian Statistics {\em per se} and for comprehensive
coverage, address the reader to \cite{marin:robert:2007} and \cite{robert:2007}, (again) among others.

\section{Bayesian computational challenges}\label{sec:bayesiancomputation:chall}
Bayesian Statistics being a complete inferential methodology, its scope encompasses
the whole range of standard statistician inference (and design), 
from point estimation to testing, to\index{Statistics!nonparametric} 
model selection, and to non-parametrics. In principle, once a prior distribution
has been chosen on the proper space, the whole inferential machinery is set and
the computation of estimators is usually automatically derived from this setup.
Obviously, the practical or numerical derivation of these procedures may be
exceedingly difficult or even impossible, as we will see in a few selected
examples. Before, we proceed with an incomplete typology of the categories
and difficulties met by Bayesian inference. First, let us point out that
computational difficulties may originate from one or several of the following
items:
\begin{enumerate}
\renewcommand{\theenumi}{(\roman{enumi})}
\item use of a complex parameter space, as for instance in constrained\index{parameter}
parameter\index{parameter space!constrained}\index{space!parameter}\index{parameter space}
sets like those resulting from imposing stationarity constraints in dynamic models;
\item use of a complex sampling model with an intractable likelihood, as for instance in
missing data and graphical models;\index{prior distribution}
\item use of a huge dataset;\index{dataset!large}\index{likelihood!intractable}
\item use of a complex prior distribution (which may be the posterior distribution
associated with an earlier sample);\index{prior distribution!selection of a}
\item use of a complex inferential procedure.
\end{enumerate}

\subsection{Bayesian point estimation}\label{sub:bayesian:posti}
In a formalised representation of Bayesian inference, the statistician
is given (or selects) a triplet\index{Decision theory}
\begin{itemize}
\item a sampling distribution, $f(x|\theta)$, usually associated with an
observation (or a sample) $x$;
\item a prior distribution $\pi(\theta)$, defined on the parameter space $\Theta$;\index{parameter space}
\item a loss function  $\mathrm{L}(\theta,d)$ that compares the decisions (or estimations)
$d$ for the true value $\theta$ of the parameter.\index{space!parameter}
\end{itemize}
Using $(f,\pi,\mathrm{L})$ and an observation $x$, 
the Bayesian inference is {\em always} given as the solution to the minimisation programme
$$
\min_d\,\int_\Theta L(\theta,d)\,f(x|\theta)\,\pi(\theta)\,\text{d}\theta\,,
$$
equivalent to the minimisation programme
$$
\min_d\,\int_\Theta L(\theta,d)\,\pi(\theta|x)\,\text{d}\theta\,.
$$
The corresponding procedure is thus associated, for every $x$,
to the solution of the above programme \citep[see, e.g.][Chap.~2]{robert:2007}.

There are therefore two levels of computational difficulties with this resolution:
first the above integral must be computed. Second, it must be minimised in $d$. For
the most standard losses, like the traditional squared error loss,
$$
\mathrm{L}(\theta,d) = \left\vert \theta -d\right\vert^2,
$$
the solution to the minimisation problem is universally\footnote{In this chapter,
the denomination {\em universal} is used in the sense of
{\em uniformly over all distributions.}} known. For instance, for the
squared error loss, it is the posterior mean,
$$
\int_\Theta \theta \,\pi(\theta|x)\,\text{d}\theta =
\int_\Theta \theta \,f(x|\theta)\,\pi(\theta)\,\text{d}\theta \bigg/
\int_\Theta f(x|\theta)\,\pi(\theta)\,\text{d}\theta \,,
$$
which still requires the computation of both integrals and thus whose complexity
depends on the complexity of $\Theta$, $f(x|\theta)$, and $\pi(\theta)$.

\begin{exemple}\label{ex:nor1} For a normal distribution $\mathcal{N}(\theta,1)$, 
the use of a\index{prior distribution!conjugate}
so-called conjugate prior \citep[see, e.g.,][Chap. 3]{robert:2007} 
$$
\theta\sim\mathcal{N}(\mu,\epsilon)\,,
$$
leads to a closed form expression for the mean, since
\begin{align*}
\int_\Theta &\theta \,f(x|\theta)\,\pi(\theta)\,\text{d}\theta \bigg/
\int_\Theta f(x|\theta)\,\pi(\theta) \,\text{d}\theta =\\ 
\int_\mathbb{R} &\theta \exp\frac{1}{2}\left\{ 
-\theta^2(1+\epsilon^{-2}) +2\theta(x
+\epsilon^{-2}\mu)\right\} \text{d}\theta \\
&\qquad \bigg/
\int_\mathbb{R} \exp\frac{1}{2}\left\{ -\theta^2(1+\epsilon^{-2})
+2\theta(x+\epsilon^{-2}\mu)\right\} \text{d}\theta =
\frac{x+\epsilon^{-2}\mu}{1+\epsilon^{-2}}\,.
\end{align*}
On the other hand, if we use instead a more involved prior distribution
like a poly-$t$ distribution \citep{bauwens:1985},\index{poly-$t$ distribution}
$$
\pi(\theta) = \prod_{i=1}^k \left[ \alpha_i+(\theta-\beta_i)^2 \right]^{-\nu_i} 
\qquad \alpha,\nu>0
$$
the above integrals cannot be computed in closed form anymore. This is {\em not}
a toy example in that the problem may occur after a sequence of $k$ Student's $t$ 
observations, or with a sequence of normal observations whose variance is unknown.
\end{exemple}\index{tdistribution@{$t$-distribution}}

The above example is one-dimensional, but, obviously, bigger challenges await the
Bayesian statistician when she wants to tackle high-dimensional 
problems.\index{dimension!high}

\begin{exemple}\label{ex:glm} In a generalised linear model, a conditional distribution of
$y\in\mathbb{R}$ given $x\in\mathbb{R}^p$ is defined via a density from
an exponential family\index{exponential family}\index{model!generalised linear}
$$
y|x\sim\exp\left\{ y\cdot\theta(x) - \psi(\theta(x)) \right\}
$$
whose natural parameter $\theta(x)$ depends on the conditioning variable $x$,\index{variable!conditioning}
$$
\theta(x) = g(\beta^\Tee x)\,,\qquad\beta\in\mathbb{R}^p
$$
that is, linearly modulo the transform $g$. Obviously, in practical applications
like Econometrics, $p$ can be quite large. Inference on $\beta$ (which is the true
parameter of the model) proceeds through the 
posterior distribution (where\index{posterior distribution}
$\mathbf{x}=(x_1,\ldots,x_T)$ and $\mathbf{y}=(y_1,\ldots,y_T)$)
\begin{align}
\pi(\beta|\mathbf{x},\mathbf{y}) &\propto
\prod_{t=1}^T \exp\left\{ y_t\cdot\theta(x_t) 
	- \psi(\theta(x_t)) \right\} \,\pi(\beta)\nonumber\\
&= \exp\left\{ \sum_{t=1}^T y_t\cdot\theta(x_t) 
	- \sum_{t=1}^T \psi(\theta(x_t)) \right\} \,\pi(\beta) \,,\nonumber
\end{align}
which rarely is available in closed form. In addition, in some cases $\psi$ may
be costly simply to compute and in others $T$ may be large or even very large. Take
for instance the case of the dataset processed by \cite{abowd:kramarz:margolis:1999}, which
covers twenty years of employment histories for over a million workers, with $x$ including
indicator variables for over one hundred thousand companies.\index{variable!indicator}

Complexity starts sharply increasing if we introduce in addition random effects to the model, 
that is writing $\theta(x)$ as $g(\beta^\Tee x+\epsilon(x))$, where $\epsilon(x)$ is a random
perturbation indexed by $x$. For instance, in the above employment dataset, it may correspond
to a worker effect or to a company effect. The difficulty is that those random variables can
very rarely be integrated out into a closed-form marginal distribution. They must therefore be
included within the model parameter, which then increases its dimension severalfold.\index{dataset!employment}
\end{exemple}

A related, although conceptually different, inferential issue concentrates 
upon {\em prediction},\index{prediction}
that is, the approximation of a distribution related with the parameter of interest, 
say $g(y|\theta)$, based on the observation of $x\sim f(x|\theta)$. The {\em 
predictive distribution} is then defined as\index{distribution!predictive}\index{parameter!of 
interest}
$$
\pi(y|x) = \int_\Theta g(y|\theta) \pi(\theta|x) \text{d}\theta\,.
$$
A first difference with the standard point estimation perspective is obviously that the parameter
$\theta$ vanishes through the integration. A second and more profound difference is that this
parameter is not necessarily well-defined anymore. As will become clearer in a following Section,
this is a paramount feature in setups where the model is not well-defined and where the statistician
hesitates between several (or even an infinity of) models. It is also a case where the standard
notion of identifiability\index{identifiability} 
is irrelevant, which paradoxically is a ''plus" from the computational
point of view, as seen below in, e.g., Example \ref{ex:glmPbit2}.

\begin{exemple}\label{ex:AR}
Recall that an $AR(p)$ model is given as the\index{AR model}
{\em auto-regressive} representation of a time series,\index{model!AR}
$$
x_t = \sum_{i=1}^p \theta_{i} x_{t-i} + \sigma\varepsilon_t \;.
$$ 
It is often the case that the order $p$ of the $AR$ model is not fixed {\em a
priori}, but has to be determined from the data itself. Several models are
then competing for the ``best" fit of the data, but if the prediction of the\index{fit!data}
next value $x_{t+1}$ is the most important part of the inference, the order\index{AR model!order}
$p$ chosen for the best fit is not really relevant. Therefore, all models can
be considered in parallel and aggregated through the predictive distribution
$$
\pi(x_{t+1}|x_t,\ldots,x_1) \propto \int f(x_{t+1}|x_t,\ldots,x_{t-p+1}) \pi(\theta,p|
x_t,\ldots,x_1) \text{d}p\,\text{d}\theta\,,
$$
which thus amounts to integrating over the parameters of all models,
simultaneously:
$$
\sum_{p=0}^\infty
	\int f(x_{t+1}|x_t,\ldots,x_{t-p+1}) \pi(\theta|p,x_t,\ldots,x_1) \,\text{d}\theta
	\,\pi(p|x_t,\ldots,x_1)\,.
$$
Note the multiple layers of complexity in this case:
\begin{enumerate}
\renewcommand{\theenumi}{(\roman{enumi})}
\item if the prior distribution on $p$ has an infinite support, the
integral simultaneously considers an infinity of models, with parameters
of unbounded dimensions;\index{dimension!unbounded}
\item the parameter $\theta$ varies from model $AR(p)$ to model $AR(p+1)$,
so must be evaluated differently from one model to another. In particular, if the
stationarity constraint usually imposed in these models is taken into account, the
constraint on $(\theta_1,\ldots,\theta_p)$ varies\footnote{To impose the stationarity constraint
when the order of the $AR(p)$ model varies,\index{stationarity}
it is necessary to reparameterise this model in terms of either 
the partial autocorrelations or of the\index{autocorrelation!partial}
roots of the associated lag polynomial. \citep[See, e.g.,][Section 4.5.]{robert:2007}}
between model $AR(p)$ and model 
$AR(p+1)$;
\item prediction is usually used sequentially: every tick/second/hour/day,
the next value is predicted based on the past values $x_t,\ldots,x_1)$. Therefore
when $t$ moves to $t+1$, the entire posterior distribution $\pi(\theta,p|
x_t,\ldots,x_1)$ must be re-evaluated again, possibly with a very tight time constraint
as for instance in financial or radar tracking applications.\index{prediction!sequential}
\end{enumerate}
\end{exemple}

We will discuss this important problem in deeper details after the testing
section, as part of the model selection problematic.\index{testing 
hypotheses}\index{hypotheses}

\subsection{Testing hypotheses}\label{sub:bayesian:test}
A domain where both the philosophy and the implementation of
Bayesian inference are at complete odds with the classical approach
is the area of testing of hypotheses. At a primary level, this is obvious
when opposing the Bayesian evaluation of an hypothesis $H_0:\theta\in\Theta_0$
$$
\hbox{Pr}^\pi (\theta\in\Theta_0|x)
$$
with a Neyman--Pearson $p$-value\index{Neyman--Pearson theory}
$$
\sup_{\theta\in\Theta_0}\,\mbox{Pr}_{\theta}(T(X)\ge T(x))
$$
where $T$ is an appropriate statistic, with observed value $T(x)$. The first quantity 
involves an integral over the {\em parameter} space, while the second provides an\index{space!parameter}
evaluation over the {\em observational} space. At a secondary level, the two answers
may also strongly disagree even when the number of observations goes to infinity, although
there exist cases and priors for which they agree to the order \smash{$\mathrm{O}(n^{-1})$} 
or even \smash{$\mathrm{O}(n^{-3/2})$}.\index{pvalue@{$p$-value}}
\citep[See][Section 3.5.5 and Chapter 5, for more details.]{robert:2007}

From a computational point of view, most Bayesian evaluations of the relevance of an hypothesis---also
called the {\em evidence}---given a sample $x$ involve marginal distributions
\begin{equation}\label{eq:mardi}
\int_{\Theta_i} f(x|\theta_i) \pi_i(\theta_i)\,\text{d}\theta_i
\end{equation}
where $\Theta_i$ and $\pi_i$ denote the parameter space and the corresponding prior, respectively,
under hypothesis $H_i$ $(i=0,1)$. For instance, the {\it Bayes factor}\index{Bayes factor} 
is defined as the ratio\index{ratio!of posterior probabilities}\index{integral!ratio}
of the posterior probabilities of the null and the alternative hypotheses over the ratio 
of the prior probabilities of the null and the alternative hypotheses, i.e.,
$$
B^\pi_{01}(x) = \dfrac{\text{Pr}(\theta \in \Theta_ 0\mid x) }{ \text{Pr}(\theta \in \Theta_1\mid x)}
\bigg/  \dfrac{\pi(\theta \in \Theta_ 0) }{ \pi(\theta \in \Theta_ 1)}.
$$
This quantity is instrumental in the computation of the posterior probability
$$
\text{Pr}(\theta \in \Theta_ 0\mid x) = \frac{1}{1+B^\pi_{10}(x)}
$$
under equal prior probabilities for both $\Theta_0$ and $\Theta_1$. It is also the central
tool in practical (as opposed to decisional) Bayesian testing \citep{Jeffreys:1961} and can
be seen as the Bayesian equivalent of the likelihood ratio.\index{likelihood!ratio}

The first ratio in $B^\pi_{01}(x)$ is then the ratio of integrals of the form (\ref{eq:mardi})
and it is rather common to face difficulties in the computation of {\em both} 
integrals.\footnote{In this presentation of Bayes factors, we completely bypass the methodological
difficulty of defining $\pi(\theta \in \Theta_0)$ when $\Theta_0$ is of measure $0$
for the original prior $\pi$ and refer the reader to Robert (2007, Section 5.2.3) 
and Marin and Robert (2007, Section 2.3.2) for proper coverage of this issue.}

\begin{exemple}[Continuation of Example \ref{ex:glm}]\label{ex:glm2} 
In the case of the generalised linear
model, a standard testing situation is to decide whether or not a factor, $x_1$ say, 
is influential on the dependent variable $y$. This is often translated as testing whether or not
the corresponding component of $\beta$, $\beta_1$, is {\em equal} to $0$, i.e.~$\Theta_0=\{\beta;
\beta_1=0\}$. If we denote by $\beta_{-1}$ the {\em other} components of $\beta$, the Bayes factor
for this hypothesis will be
\begin{align}
\int_{\mathbb{R}^p} \exp&\left\{ \sum_{t=1}^T y_t\cdot g(\beta^\Tee x_t) 
	- \sum_{t=1}^T \psi(g(\beta^\Tee x_t)) \right\} \,\pi(\beta) \,\text{d}\beta\bigg/\nonumber\\
\int_{\mathbb{R}^{p-1}} &\exp\left\{ \sum_{t=1}^T y_t\cdot g(\beta^\Tee_{-1}(x_t)_{-1}) 
 -\sum_{t=1}^T \psi( \beta^\Tee_{-1}(x_t)_{-1} ) \right\} \,\pi_{-1}(\beta_{-1} ) 
 \,\text{d}\beta_{-1} \,,\nonumber
\end{align}
when $\pi_{-1}$ is the prior constructed for the null hypothesis and when the prior weights of
$H_0$ and of the alternative are both equal to $1/2$. Obviously, besides the normal conjugate
case, both integrals cannot be computed in a closed form.
\end{exemple}

In a related manner, {\em confidence regions} are also mostly intractable, being defined through
the solution to an implicit equation. Indeed, the Bayesian confidence region for a parameter
$\theta$ is defined as the {\em highest posterior region},\index{confidence!regions}\index{highest 
posterior region}
\begin{equation}\label{eq:implicon}
\left\{ \theta; \pi(\theta|x) \ge k(x) \right\}\,
\end{equation}
where $k(x)$ is determined by the coverage constraint
$$
\hbox{Pr}^\pi(\pi(\theta|x) \ge k(x) | x ) = \alpha\,,
$$
$\alpha$ being the confidence level. While the normalising constant is not necessary to
construct a confidence region, the resolution of the implicit equation (\ref{eq:implicon})
is rarely straightforward!\index{confidence!level} However, simulation-based equivalents generally
produce acceptable approximations in a straightforward manner when both the prior and the likelihood
can be numerically computed.

\begin{exemple} Consider a binomial observation $x\sim\mathcal{B}(n,\theta)$ with a conjugate
prior distribution, $\theta\sim\mathcal{B}e(\gamma_1,\gamma_2)$. In this case, the posterior
distribution is available in closed form,\index{model!binomial}\index{distribution!binomial}
$$
\theta|x \sim \mathcal{B}e(\gamma_1+x,\gamma_2+n-x)\,.
$$
However, the determination of the $\theta$'s such that
$$
\theta^{\gamma_1+x-1}(1-\theta)^{\gamma_2+n-x-1} \ge k(x) 
$$
with 
$$
\hbox{Pr}^\pi\left(
\theta^{\gamma_1+x-1}(1-\theta)^{\gamma_2+n-x-1} \ge k(x) | x \right) = \alpha
$$
is not possible analytically. It actually implies two levels of numerical difficulties:
\begin{enumerate}
\renewcommand\theenumi{\arabic{enumi}.}
\item find the solution(s) to $\theta^{\gamma_1+x-1}(1-\theta)^{\gamma_2+n-x-1}=k$,
\item find the $k$ corresponding to the right coverage,
\end{enumerate}
and each value of $k$ examined in step 2. requires a new resolution of step 1. However, 
$k$ can also be interpreted as the $(1-\alpha)$ quantile of the random variable 
$\theta^{\gamma_1+x-1}(1-\theta)^{\gamma_2+n-x-1}$, hence derived from a large sample
of $\theta$'s in a Monte Carlo perspective.
\end{exemple}

The setting is usually much more complex when $\theta$ is a multidimensional parameter,
because the interest is usually in getting marginal confidence sets. Example \ref{ex:glm}
is an illustration of this setting: deriving a confidence region on one component,
$\beta_1$ say, first involves computing the marginal posterior distribution of this
component. As in Example \ref{ex:glm2}, the integral
$$
\int_{\mathbb{R}^{p-1}} \exp\left\{ \sum_{t=1}^T y_t\cdot g(\beta^\Tee x_t) 
 -\sum_{t=1}^T \psi( \beta^\Tee x_t ) \right\} \,\pi_{-1}(\beta_{-1} )\,\text{d}\beta_{-1} \,,
$$
which is proportional to $\pi(\beta_1|x)$, is most often intractable. Fortunately, the simulation
approximations mentioned above are also available to bypass this integral computation.

\subsection{Model choice}\label{sub:bayesian:mocho}
Although they are essentially identical from a conceptual viewpoint, we 
do distinguish here between {\em model choice}\index{model choice} 
and testing, partly because the former leads to
further computational difficulties, and partly because it encompasses a larger
scope of inferential goals than mere testing. Note first that model choice
has been the subject of considerable effort in the past decades, and has seen
many advances, including the coverage of problems of higher complexity and the
introduction of new concepts. We stress that such advances
mostly owe to the introduction of new computational methods.

As discussed in further details in Robert (2007, Chapter 7), 
the inferential action related with model choice does take place on a wider scale
than simple testing: 
it covers and compares models, rather than parameters, which makes 
the sampling distribution $f(x)$ ``more unknown"
than simply depending on an undetermined parameter. In some respect, it is
thus closer to estimation than to regular testing\index{model choice!and
testing}.  In any case, it requires a more precise evaluation of the
consequences of choosing the ``wrong" model or, equivalently of deciding
which model is the most appropriate for the data at hand. It is thus both
broader and less definitive than deciding whether $H_0:\,\theta_1=0$ is true.
At last, the larger inferential scope mentioned in the first point means 
that we are leaving for a while the well-charted domain of solid parametric models.

From a computational point of view, model choice involves more complex 
structures that, almost systematically, require advanced tools, like
simulation methods which can handle collections of parameter spaces (also called
{\it spaces of varying dimensions}\index{space!varying dimension}),
specially designed for model comparison.

\begin{exemple}\label{ex:mix}
A mixture of distributions is the representation of a distribution (density)
as the weighted sum of standard distributions (densities). For instance, a 
mixture of Poisson distributions, denoted as\index{mixture!Poisson distributions} 
$$
\sum_{i=1}^k p_i{\mathcal P}(\lambda_i)\,
$$
has the following density:
$$
\hbox{Pr}(X=k) = \sum_{i=1}^k p_i\,\frac{\lambda_i^k}{k!}\,e^{-\lambda_i}\,.
$$
This representation of distributions is multi-faceted and can be used in
populations with known heterogeneities (in which case a component of the mixture
corresponds to an homogeneous part of the population) as well as a
non-parametric modelling of unknown populations. This means that, in some cases,
$k$ is known and, in others, it is both unknown and part of the inferential problem.
\end{exemple}\index{heterogeneous populations}

First, consider the setting where several (parametric) models are in competition,
$$
{\mathfrak M}_i : x \sim f_i(x|\theta_i) , \qquad \theta_i\in\Theta_i,\quad i \in I\,,
$$
the index set $I$ being possibly infinite.  From a Bayesian point of view, 
a prior distribution must be constructed for each model $\mathfrak M_i$ as if it were the only 
and true model under consideration since, in most perspectives except {\em model averaging},\index{model 
averaging} {\em only one} of these models will be selected and used as the only\index{parameter space}
and true model.  The parameter space\index{model choice!parameter space}\index{space!parameter}
associated with the above set of models can be written as
\begin{equation}\label{eq:bigset}
  {\mathbf \Theta} = \bigcup_{i\in I} \{i\}\times\Theta_i \,,
\end{equation}
the model indicator $\mu\in I$ being now part of the parameters.
So, if the modeller allocates probabilities $p_i$ to the indicator values,  
that is, to the models ${\mathfrak M}_i$
$(i\in I)$, and if she then defines priors $\pi_i(\theta_i)$ on the parameter  subspaces
$\Theta_i$, things fold over by virtue of Bayes's theorem\index{Bayes!theorem}, since one can compute
$$
p({\mathfrak M}_i|x) = \text{Pr}(\mu = i | x) = \dfrac{ \displaystyle{p_i 
\int_{\Theta_i} f_i(x|\theta_i)
\pi_i(\theta_i) \text{d}\theta_i}
}{ 
\displaystyle{\sum_j p_j \int_{\Theta_j} f_j(x|\theta_j) 
\pi_j(\theta_j) 
\text{d}\theta_j}  }  \,.
$$
While a common solution based on this prior modelling is simply
to take the (marginal) MAP estimator\index{estimator!maximum
a posteriori (MAP)}  of $\mu$, that
is, to determine the model with the largest $p({\mathfrak M}_i|x)$, or even
to use directly the average\index{model!averaging}
$$
\displaystyle{ \sum_j p_j \int_{\Theta_j} f_j(y|\theta_j)  \pi_j(\theta_j|x)
\text{d}\theta_j } = \sum_j p({\mathfrak M}_j|x) \, m_j(y) 
$$
as a predictive density in $y$ in {\em model averaging}, 
a deeper-decision theoretic evaluation is often necessary.

\begin{exemple}[Continuation of Example \ref{ex:AR}]\label{ex:AR2}
In the setting of the $AR(p)$ models, when the order $p$ of the dependence 
is unknown, model averaging as presented in Example \ref{ex:AR}
is not always a relevant solution when the statistician wants to estimate 
this order $p$ for different purposes. Estimation is then a more appropriate
perspective than testing, even though care must be taken because of the
discrete nature of $p$. (For instance, the posterior expectation of $p$ is
not an appropriate estimator!)\index{AR model}\index{model!AR}
\end{exemple}\index{estimation vs.~testing}


As stressed earlier in this Section, the computation of predictive densities,  
marginals, Bayes factors,\index{Bayes factor!computation} 
and other quantities related to the model choice procedures is
generally very involved, with specificities that call for tailor-made solutions: 
\begin{itemize}
\item[--]  The computation of integrals is increased by a factor corresponding
	    to the number of models under consideration.
\item[--]  Some parameter spaces are infinite-dimensional, as in non-parametric
	     settings and that may cause measure-theoretic complications.\index{dimension!unbounded}
\item[--]  The computation of posterior or predictive quantities
	     involves integration over different parameter spaces and
	     thus increases the computational burden, since
	     there is no time savings from one subspace to the next.
\item[--]  In some settings, the size of the collection
	     of models is very large or even infinite and\index{infinite collection of models}
	     some models cannot be explored. For instance, in Example \ref{ex:glm2},
	     the collection of all submodels is of size $2^p$ and some pruning method
	     must be found in variable selection to avoid exploring the whole tree of
	     all submodels.\index{variable!selection}
\end{itemize}\index{Bayesian!Statistics|)} 

\section{Monte Carlo Methods}\label{sec:moca}
The natural approach to these computational problems is to use computer simulation and
Monte Carlo techniques, rather than numerical methods, simply because there is much
more to gain from exploiting the probabilistic properties of the integrands rather than
their analytical properties. In addition, the dimension of most problems considered in
current Bayesian Statistics is such that very involved numerical methods should be used
to provide a satisfactory approximation in such integration or optimisation problems.
Indeed, down-the-shelf numerical methods cannot handle integrals in moderate dimensions 
and more advanced numerical integration methods require analytical studies on the
distribution of interest.\index{Monte Carlo techniques|(}\index{simulation}

\subsection{Preamble: Monte Carlo importance sampling}
Given the statistical nature of the problem, the approximation of an integral
like\index{integral!approximation}
$$
\mathfrak{I} = \int_\Theta h(\theta) f(x|\theta) \pi(\theta) \,\text{d}\theta,
$$
should indeed take advantage of the special nature of\index{simulation}
$\mathfrak{I}$, namely, the fact that $\pi$ is a probability 
density\footnote{The prior distribution can be used for importance sampling
only if it is a proper prior and not a $\sigma$-finite
measure.} or, instead, that $f(x|\theta) \pi(\theta)$ is proportional
to a density.  As detailed in Chapter II.2
this volume, or in Robert and Casella (2004, Chapter 3)\index{prior distribution!proper}
and \cite{robert:casella:2010}\index{Monte Carlo techniques!approximation},
the {\it Monte Carlo method} was introduced by \cite{Metropolis:Ulam:1949} for 
this purpose. For instance, if it is possible to generate (via a computer)
random variables $\theta_1, \ldots,\theta_m$ from $\pi(\theta)$, the average
$$
\dfrac{1}{m}\sum_{i=1}^m h(\theta_i) f(x|\theta_i) 
$$
converges (almost surely) to $\mathfrak{I}$ when $m$ goes to $+\infty$, 
according to the Law of Large Numbers. Obviously, if an i.i.d.\ sample of $\theta_i$'s
from the posterior distribution $\pi(\theta|x)$ can be produced, the average
\begin{equation}\label{eq:oneaverage}
	\dfrac{1}{m}\sum_{i=1}^m h(\theta_i)
\end{equation}
converges to\index{Law of Large Numbers}
$$
\mathbb{E}^\pi [h(\theta)|x] = \frac{\int_\Theta h(\theta) f(x|\theta) 
  \pi(\theta) \,\text{d}\theta} {\int_\Theta f(x|\theta) \pi(\theta) \,\text{d}\theta}
$$
and it usually is more interesting to use this approximation, rather than
$$
\sum_{i=1}^m h(\theta_i) f(x|\theta_i)  \bigg/
\sum_{i=1}^m f(x|\theta_i)
$$
when the $\theta_i$'s are generated from $\pi(\theta)$, especially when $\pi(\theta)$
is flat compared with $\pi(\theta|x)$.

In addition, if the posterior variance $\mathrm{var}(h(\theta)|x)$ is finite, 
the Central Limit Theorem applies to the empirical average (\ref{eq:oneaverage}), 
which is then asymptotically normal with variance $\mathrm{var}(h(\theta)|x)/m$.
Confidence regions can then be built from this normal approximation and, most
importantly, the magnitude of the error remains of order $1/\sqrt m$, whatever the dimension
of the problem, in opposition with numerical methods.\footnote{The constant order
of the Monte Carlo error does not imply that
the computational effort remains the same as the dimension increases, most obviously, but
rather that the decrease (with $m$) in variation has the rate $1/\sqrt m$.} \citep[See
also][Chapter 4, for more details on the convergence assessment based on the
CLT.]{robert:casella:2004,robert:casella:2009}\index{Central Limit Theorem}\index{convergence 
assessment}\index{computational effort}

The Monte Carlo method actually applies in a much wider generality than the above
simulation from $\pi$. For instance, because $\mathfrak{I}$ can be represented in an
infinity of ways as an expectation, there is no need to simulate from the distributions 
$\pi(\cdot|x)$ or $\pi$ to get a good approximation of $\mathfrak{I}$. 
Indeed, if $g$ is a probability density with $\mathrm{supp}(g)$ including the support of
	$|h(\theta)| f(x|\theta) \pi(\theta)$,
the integral $\mathfrak{I}$  can also be represented as an expectation 
against $g$, namely
$$
  \int \frac{h(\theta) f(x|\theta) \pi(\theta)}{g(\theta)} g(\theta) \,\text{d}\theta.
$$
This representation leads to the {\it Monte Carlo method with 
importance function}\index{Monte Carlo techniques!with importance function}\index{importance function} 
$g$: generate $\theta_1,\ldots, \theta_m$ according to $g$ and approximate 
$\mathfrak{I}$ through
$$
\dfrac{1}{m} \sum_{i=1}^m h(\theta_i) \omega_i(\theta_i),
$$
with the weights $\omega(\theta_i)=f(x|\theta_i)\pi(\theta_i)/g(\theta_i)$. 
Again, by the Law of Large Numbers, this approximation almost surely converges to 
$\mathfrak{I}$. And this estimator is unbiased. In addition,
an approximation to $\mathbb{E}^\pi[h(\theta)|x]$ is given by
\begin{equation}\label{eq=9.8}
\dfrac{\sum_{i=1}^m h(\theta_i) \omega(\theta_i) }{\sum_{i=1}^m \omega(\theta_i) }.
\end{equation}
since the numerator and denominator converge to
$$
	\int_\Theta h(\theta)f(x|\theta)\pi(\theta)\, \text{d}\theta 
\qquad\hbox{and}\qquad 
	\int_\Theta f(x|\theta)\pi(\theta)\,\text{d}\theta,
$$
respectively, if $\mathrm{supp}(g)$ includes $\mathrm{supp}(f(x|\cdot)\pi)$. 
Notice that the ratio (\ref{eq=9.8}) does not depend on the normalising
constants in either $h(\theta)$, $f(x|\theta)$ or $\pi(\theta)$. The approximation
(\ref{eq=9.8}) can therefore be used in settings when some of these normalising
constants are unknown. Notice also that the {\em same} sample of $\theta_i$'s can be
used for the approximation of both the numerator and denominator
integrals: even though using an estimator in the\index{ratio!and normalising constants}
denominator creates a bias, (\ref{eq=9.8}) does converge\index{normalising constant}
to $\mathbb{E}^\pi[h(\theta)|x]$.

While this convergence is guaranteed for all densities $g$ with wide enough support,
the choice of the importance function\index{importance function!choice of} is crucial. 
First, simulation from $g$ must be easily implemented.
Moreover, the function $g(\theta)$ must be close enough to the function
$h(\theta)\pi(\theta|x)$, in order to reduce the variability of (\ref{eq=9.8}) 
as much as possible; otherwise, most of the weights $\omega(\theta_i)$ will be quite 
small and a few will be overly influential. In fact, if $\mathbb{E}^h [ h^2(\theta) 
\omega^2(\theta) ]$ is not finite, the variance of the estimator (\ref{eq=9.8}) is
infinite \citep[see][Chapter 3]{robert:casella:2004}.\index{importance function!with finite variance} 
Obviously, the dependence on $g$ of the importance function $h$ can be avoided
by proposing generic choices such as the posterior distribution $\pi(\theta|x)$.

\subsection{First illustrations}
In either point estimation or simple testing situations, the computational problem
is often expressed as a ratio of integrals.\index{ratio!of integrals}
Let us start with a toy example to set up the way Monte Carlo methods proceed and 
highlight the difficulties of applying a generic approach to the problem.

\begin{exemple}\label{ex:toy} 
Consider a $t$-distribution $\mathcal{T}(\nu,\theta,1)$
sample $(x_1,\ldots,x_n)$ with $\nu$ known. Assume in addition a flat prior
$\pi(\theta)=1$ as in a non-informative environment. While the posterior distribution
on $\theta$ can be easily plotted, up to a normalising constant (Figure \ref{fig:toy1}), 
because we are in dimension $1$,
direct simulation and computation from this posterior is impossible. 
\begin{figure}\begin{center}
\includegraphics[height=8cm,width=4cm,angle=270]{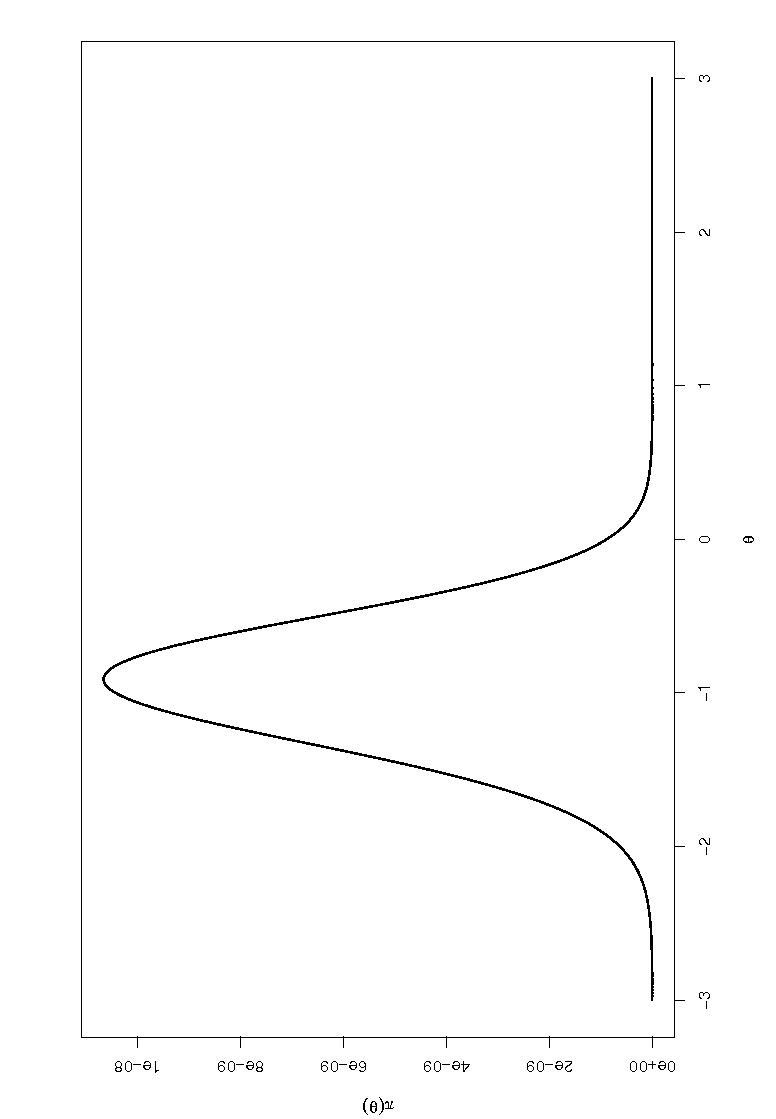}
\caption[$t$ posterior density]{\label{fig:toy1}
Posterior density of $\theta$ in the setting of Example \ref{ex:toy}
for $n=10$, with a simulated sample from $\mathcal{T}(3,0,1)$.
}
\end{center}\end{figure}

If the inferential\index{distribution!t@{$t$}}
problem is to decide about the value of $\theta$, the posterior expectation is
\begin{eqnarray*}
&&\mathbb{E}^\pi [\theta|x_1,\ldots,x_n] =
\int \theta \prod_{i=1}^n \left[ \nu + (x_i-\theta)^2 \right]^{-(\nu+1)/2} \,\text{d}\theta\\
&&\qquad \bigg/ \int \prod_{i=1}^n \left[ \nu + (x_i-\theta)^2 \right]^{-(\nu+1)/2} \,\text{d}\theta\,.
\end{eqnarray*}
This ratio of integrals is not directly computable. Since 
$(\nu + (x_i-\theta)^2)^{-(\nu+1)/2}$ is proportional to a $t$-distribution 
$\mathcal{T}(\nu,x_i,1)$ density, a solution to the approximation of the integrals
is to use {\em one} of the $i$'s to ``be" the density in both integrals. For instance,
if we generate $\theta_1,\ldots,\theta_m$ from the $\mathcal{T}(\nu,x_1,1)$ distribution,
the equivalent of (\ref{eq=9.8}) is
\begin{eqnarray}\label{eq:t_est}
\delta^\pi_m &=&\sum_{j=1}^m \theta_j  \prod_{i=2}^n \left[ \nu + (x_i-\theta_j)^2 \right]^{-(\nu+1)/2}\\
&&\qquad \bigg/ \sum_{j=1}^m \prod_{i=2}^n \left[ \nu + (x_i-\theta_j)^2 \right]^{-(\nu+1)/2}\nonumber
\end{eqnarray}
since the first term in the product has been ``used" for the simulation and the normalisation
constants have vanished in the ratio. Figure \ref{fig:toy2} is an illustration of the speed
of convergence of this estimator to the true posterior expectation: it provides the evolution
of $\delta^\pi_m$ as a function of $m$ both on average and on range 
(provided by repeated simulations of $\delta^\pi_m$). As can be seen from the graph, the average
is almost constant from the start, as it should, because of unbiasedness, 
while the range decreases very slowly, as it should, because of extreme value theory. The graph
provides in addition the $90$\%~empirical confidence interval built on 
these simulations.\footnote{The
empirical (Monte Carlo) confidence interval is not to be confused with the asymptotic 
confidence interval\index{normal approximation}\index{Monte Carlo techniques!confidence interval}
derived from the normal approximation. As discussed in Robert and Casella (2004, Chapter 4), these
two intervals may differ considerably in width, with the interval derived from the CLT being much
more optimistic!} The empirical confidence intervals are decreasing in 
$1/\sqrt{n}$, as expected from the theory. (This is further established by regressing
the log-lengths of the confidence intervals on $\log(n)$, with slope equal
to $-0.5$, as shown by Figure \ref{fig:rootN}.)
\begin{figure}\begin{center}
\includegraphics[height=8cm,width=4cm,angle=270]{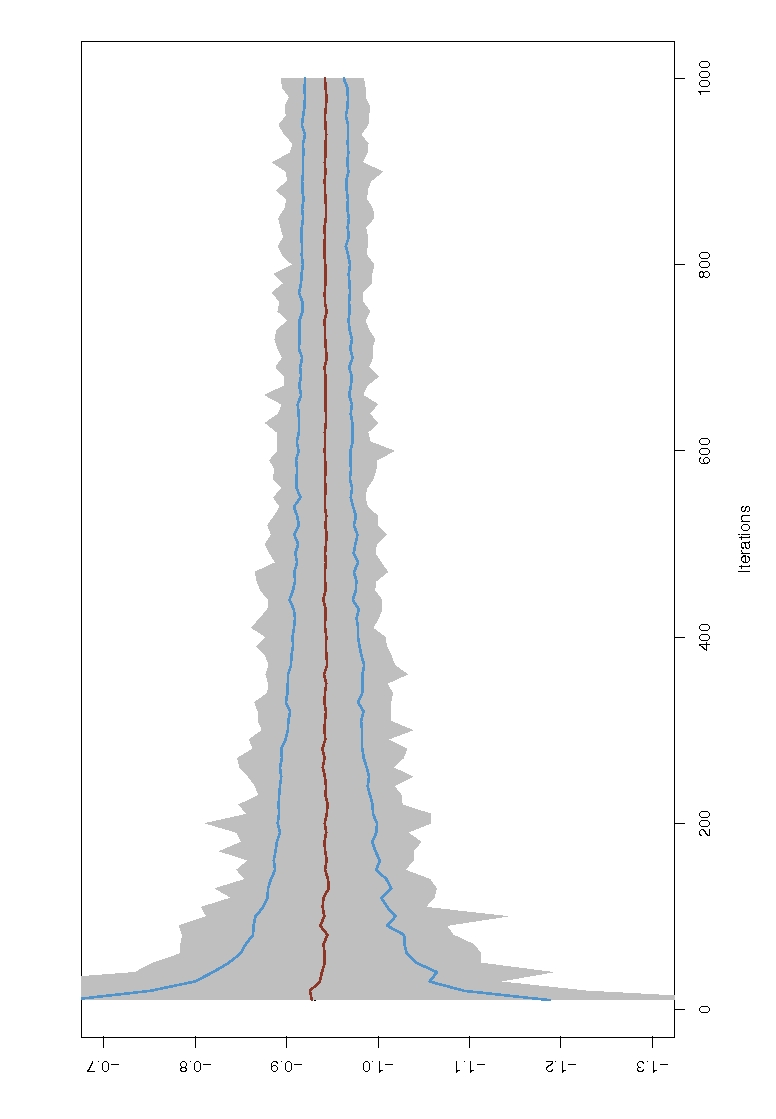}
\caption[$t$ posterior mean]{\label{fig:toy2}
Evolution of a sequence of $500$ estimators (\ref{eq:t_est}) over $1,000$ iterations:
range {\em (in gray)}, $.05$ and $.95$ quantiles, and average, obtained on the same sample
as in Figure \ref{fig:toy1} when simulating from the $t$ distribution with location $x_1$.}
\end{center}\end{figure}
\begin{figure}\begin{center}
\includegraphics[height=5cm,width=4cm,angle=270]{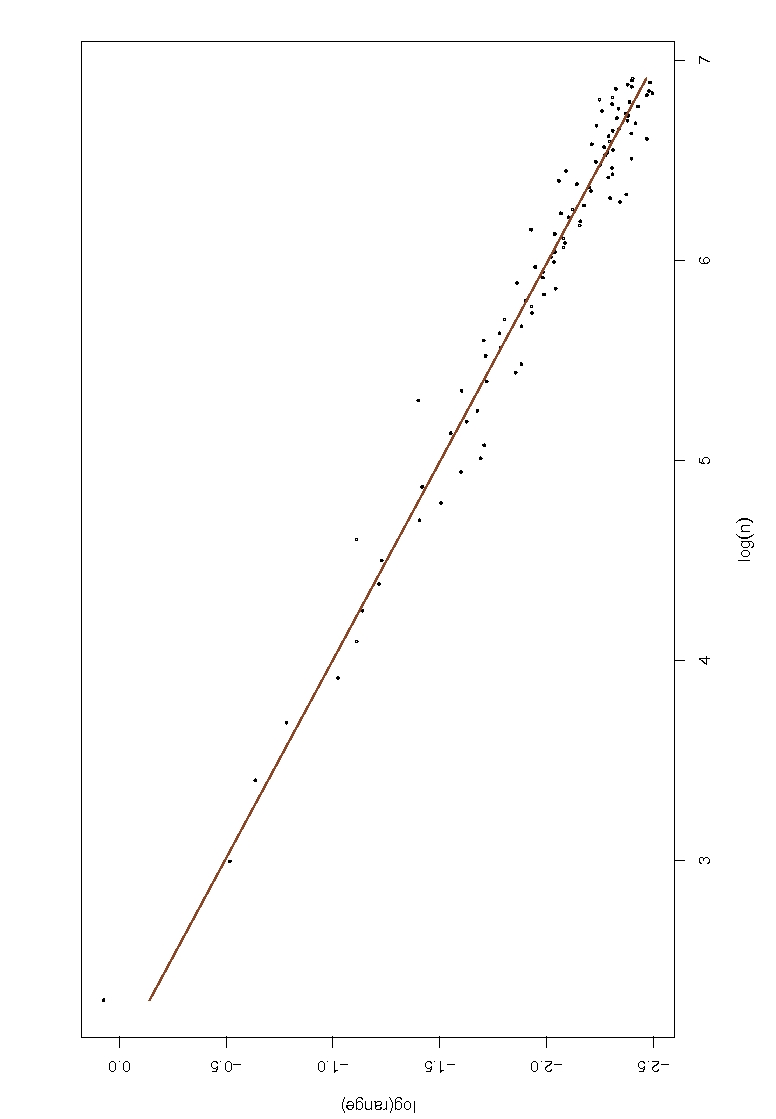}\hglue -.1cm
\caption[Range regression on $\sqrt n$]{\label{fig:rootN}
Regression of the log-ranges {\em (left)} and the log-lengths of the confidence intervals 
{\em (right)} on $\log(n)$, for the output in Figure \ref{fig:toy2}.}
\end{center}\end{figure}

Now, there is a clear arbitrariness in the choice of $x_1$ 
in the sample\index{distribution!proposal} 
$(x_1,\ldots,\allowbreak x_n)$ for the proposal $\mathcal{T}(\nu,x_1,1)$. While
any of the $x_i$'s has the same theoretical validity to ``represent" the integral and the integrating
density, the choice of $x_i$'s closer to the posterior mode (the true value of $\theta$ is $0$)
induces less variability in the estimates, as 
shown by a further simulation experiment through Figure \ref{fig:toy3}. It is fairly clear from 
this comparison that the choice of extremal values like $x_{(1)}=-3.21$ and even more
$x_{(10)}=1.72$ is detrimental to the quality of the approximation, compared with 
the median $x_{(5)}=-0.86$.
The range of the estimators is much wider for both extremes, but the influence of
this choice is also visible for the average which does not converge so quickly.\footnote{An
alternative to the simulation from one $\mathcal{T}(\nu,x_i,1)$ distribution
that does not require an extensive study on the most appropriate $x_i$ is to
use a mixture of the $\mathcal{T}(\nu,x_i,1)$ distributions. As seen in Section 
\ref{sub:MoMa:pOp}, the weights of this mixture can even be optimised automatically.}
\end{exemple}
\begin{figure}
\includegraphics[height=3.8cm,width=4cm,angle=270]{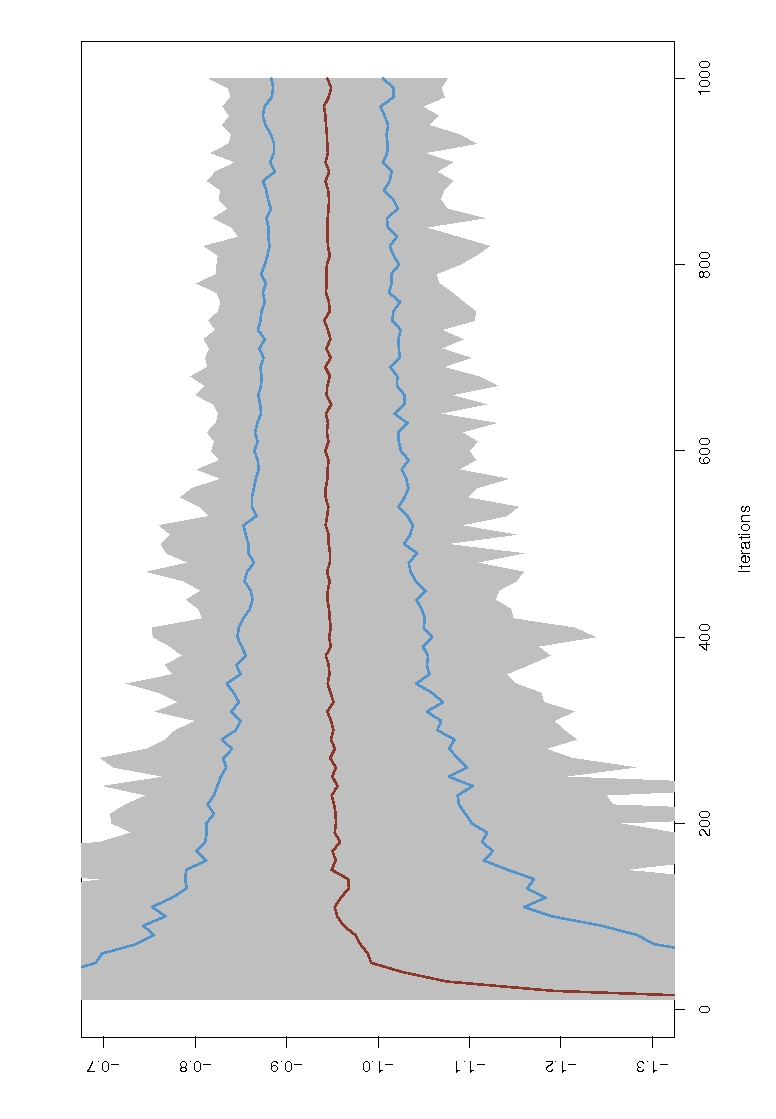}\hglue -.1cm
\includegraphics[height=3.8cm,width=4cm,angle=270]{arxivures/toy2_1.jpg} \hglue -.2cm
\includegraphics[height=3.8cm,width=4cm,angle=270]{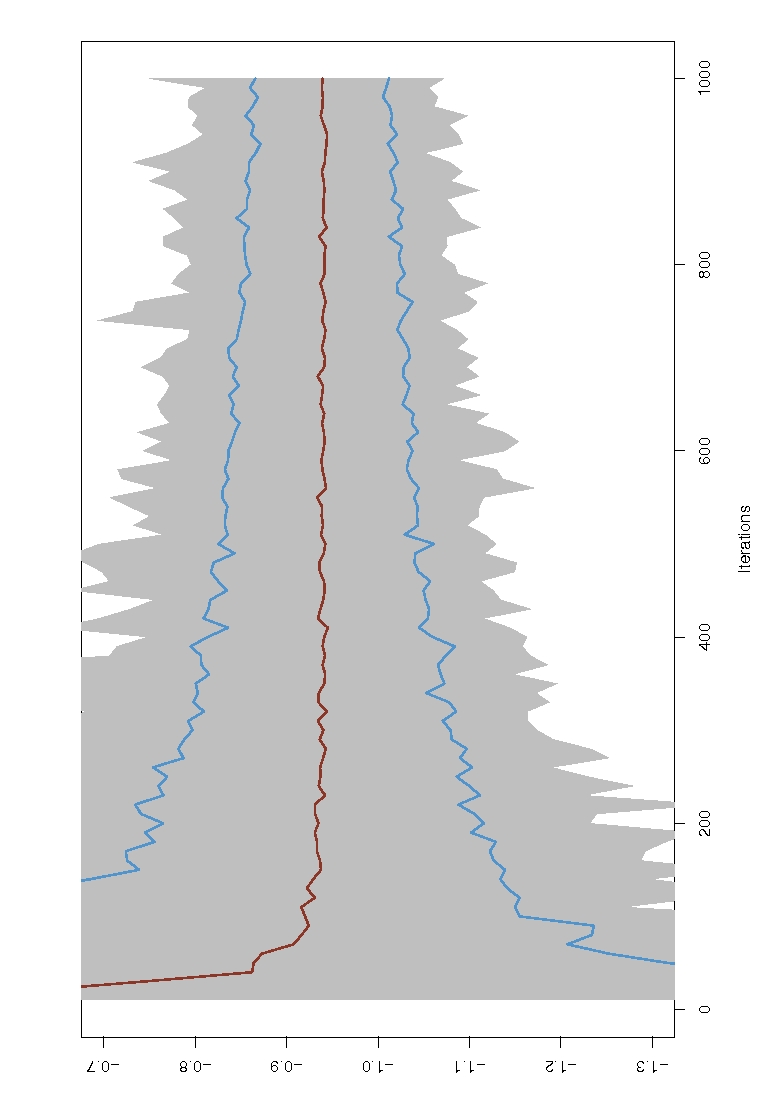}
\caption[$t$ range of ranges]{\label{fig:toy3}
Repetition of the experiment described in Figure \ref{fig:toy2} for 
three different choices of $x_i$: 
$\min x_i$, $x_{(5)}$ and $\max x_i$ {\em (from left to right)}.}
\end{figure}

This example thus shows that Monte Carlo methods, while widely available, 
may easily face inefficiency problems when the simulated values are not 
sufficiently attuned to the distribution of interest. It also demonstrates that,
fundamentally, there is no difference between importance sampling and
regular Monte Carlo, in that the integral $\mathfrak I$ can naturally
be represented in many ways.\index{Monte Carlo techniques!efficiency of}\index{importance
sampling!and regular Monte Carlo}\index{sampling!importance}

Although we do not wish to spend too much space on this issue, let us note
that the choice of the importance function gets paramount when the support
of the function of interest is {\em not} the whole space. For instance, a tail
probability, associated with $h(\theta) = \mathbb{I}_{\theta\ge\theta_0}$
say, should be estimated with an importance function whose support is
$[\theta_0,\infty)$. \citep[See][Chapter 3, for details.]{robert:casella:2004}

\begin{figure}\begin{center}
\includegraphics[height=8cm,width=4cm,angle=270]{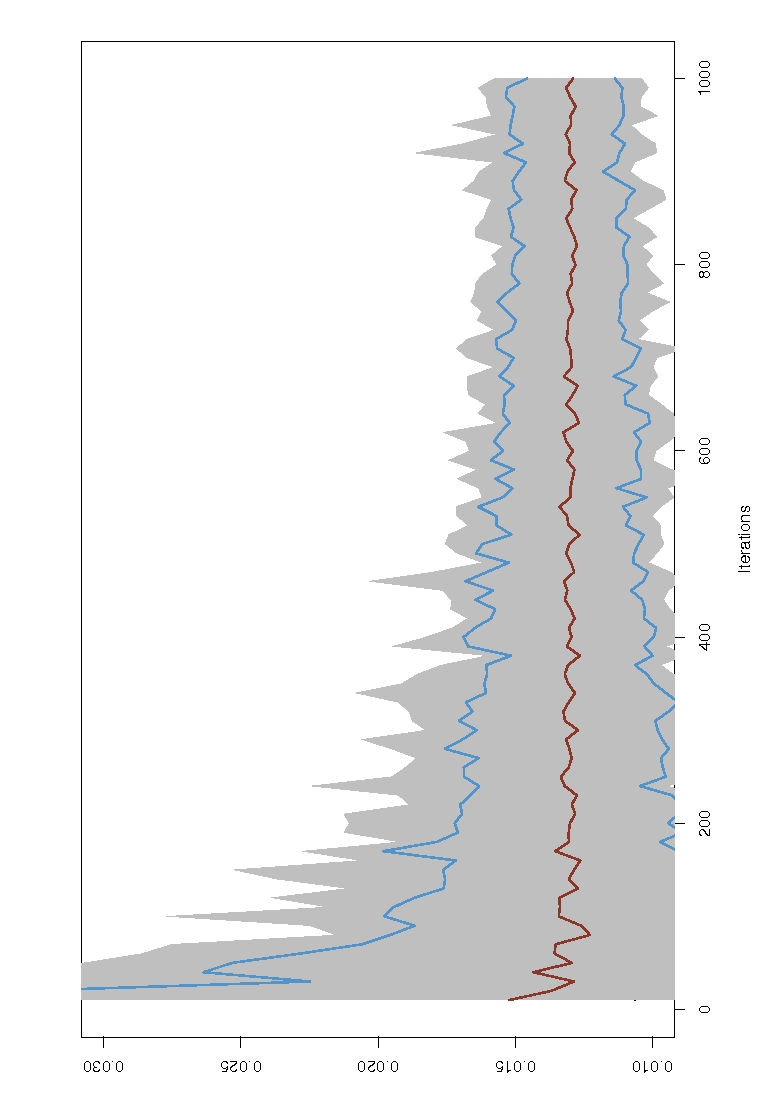}
\caption[$t$ posterior positive probability]{\label{fig:tail1}
Evolution of a sequence of $100$ estimators (\ref{eq:t_ail}) over $1,000$ iterations 
{\em (same legend as Figure \ref{fig:toy2})}.}
\end{center}\end{figure}

\begin{exemple}[Continuation of Example \ref{ex:toy}]\label{ex:toy2} 
If, instead, we wish to consider the probability that $\theta\ge 0$, using
the $t$-distribution $\mathcal{T}(\nu,x_i,1)$ is not a good idea because
negative values of $\theta$ are somehow simulated ``for nothing". A better
proposal (in terms of variance) is to use the ``folded" $t$-distribution 
$\mathcal{T}(\nu,x_i,1)$, with density proportional 
to\index{tdistribution@{$t$-distribution}!folded}\index{distribution!folded $t$}
$$
\psi_i(\theta) = \left[ \nu + (x_i-\theta)^2 \right]^{-(\nu+1)/2} +
\left[ \nu + (x_i+\theta)^2 \right]^{-(\nu+1)/2}\,,
$$
on $\mathbb{R}_+$, which can be simulated by taking the absolute value
of a $\mathcal{T}(\nu,x_i,1)$ rv. All simulated values are then positive
and the estimator of the probability is
\begin{eqnarray}\label{eq:t_ail}
\rho^\pi_m &=&\sum_{j=1}^m \prod_{i\ne k} \left[ \nu 
	+ (x_i-|\theta_j|)^2 \right]^{-(\nu+1)/2}/ \psi_k(|\theta_j|)\\
&&\qquad \bigg/ \sum_{j=1}^m \prod_{i\ne k} \left[ \nu + (x_i-\theta_j)^2 \right]^{-(\nu+1)/2}\nonumber
\end{eqnarray}
where the $\theta_j$'s are iid $\mathcal{T}(\nu,x_k,1)$. Note that this is a very special
occurrence where the {\em same} sample can be used in both the numerator and the denominator.
In fact, in most cases, two different samples have to be used, if only because the support
of the importance distribution for the numerator is {\em not} the whole space, unless, of
course, all normalising constants are known. Figure \ref{fig:tail1} reproduces earlier
figures for this problem, when using $x_{(5)}$ as the parameter of the $t$ distribution.
\end{exemple}

The above example is one-dimensional (in the parameter) and the problems exhibited there
can be found severalfold in multidimensional settings. Indeed, while Monte Carlo methods
do not suffer from the ``curse of dimension"\index{curse of dimension} 
in the sense that the error of the\index{Monte Carlo techniques!and the curse of dimension}
corresponding estimators is always decreasing in $1/\sqrt n$, notwithstanding the dimension,
it gets increasingly difficult to come up with satisfactory importance sampling distributions
as the dimension gets higher and higher. As we will see in Section \ref{sec:MoMa}, the intuition
built on MCMC methods has to be exploited to derive satisfactory importance functions.

\begin{exemple}[Continuation of Example \ref{ex:glm}]\label{ex:glmPbit}
A particular case of generalised linear model is the {\em probit 
model},\index{model!probit}
$$
\hbox{Pr}_\theta(Y=1|x) = 1 - \hbox{Pr}_\theta(Y=0|x) =
\Phi(x^\Tee\theta)\qquad\theta\in\mathbb{R}^p\,, 
$$
where $\Phi$ denotes the normal $\mathcal{N}(0,1)$ cdf. Under a flat prior
$\pi(\theta)=1$, for a sample $(x_1,y_1),\ldots,(x_n,y_n)$,
the corresponding posterior distribution is proportional to
\begin{equation}\label{eq:propos}
\prod_{i=1}^n \Phi(x_i^\Tee\theta)^{y_i} \Phi(-x_i^\Tee\theta)^{1-y_i}\,.
\end{equation}
Direct simulation from this distribution is obviously impossible since the 
computation of $\Phi(z)$ is a difficulty in itself. If we pick an importance
function for this problem, the adequation with the posterior distribution will
need to be better and better as the dimension $p$ increases. Otherwise, the repartition
of the weights will get increasingly asymmetric: very few weights will be different 
from $0$.

Figure \ref{fig:dimcomp} illustrates this degeneracy of the importance 
sampling\index{importance sampling!degeneracy of}
approach as the dimension increases. We simulate parameters $\beta$'s and datasets
$(x_i,y_i)$ $(i=1,\ldots,245)$ for dimensions $p$ ranging from $1$ to $10$, then
represented the histograms of the largest weight for $p=1,2,5,10$. The $x_i$'s
were simulated from a $\mathcal{N}_p(0,I_p)$ distribution, while the importance
sampling distribution was a $\mathcal{N}_p(0,I_p/p)$ distribution.
\begin{figure}
\includegraphics[width=4cm,height=6cm,angle=270]{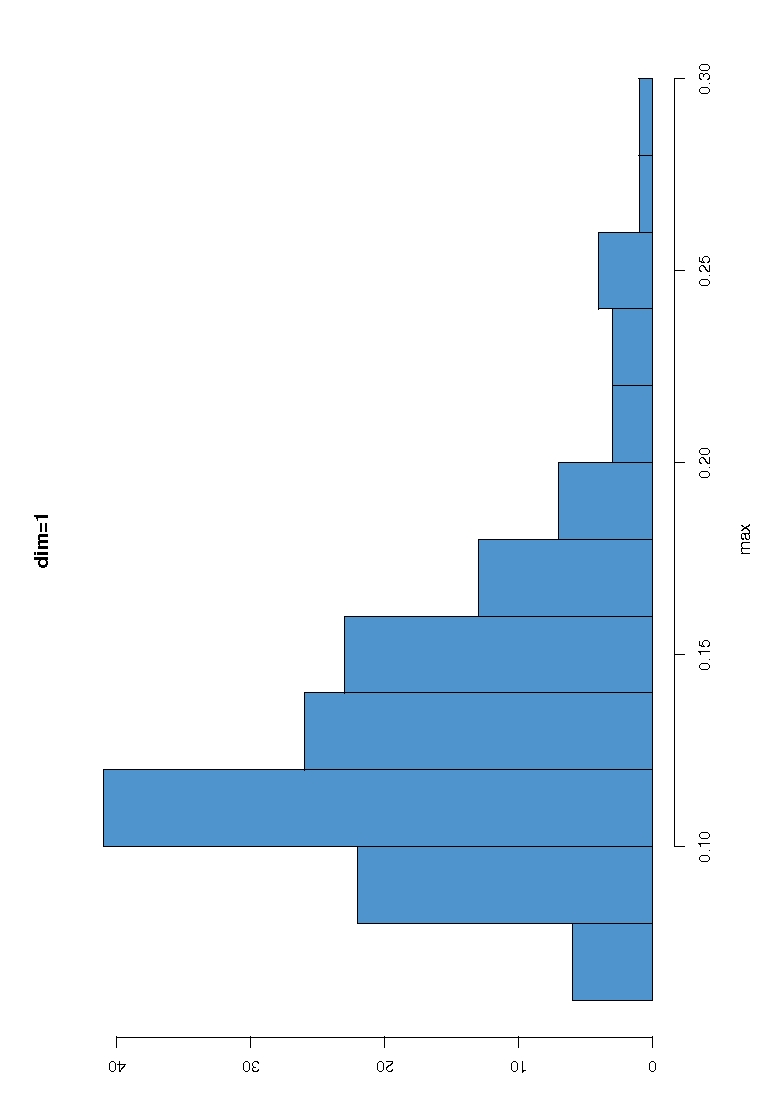}\hglue -.1cm
\includegraphics[width=4cm,height=6cm,angle=270]{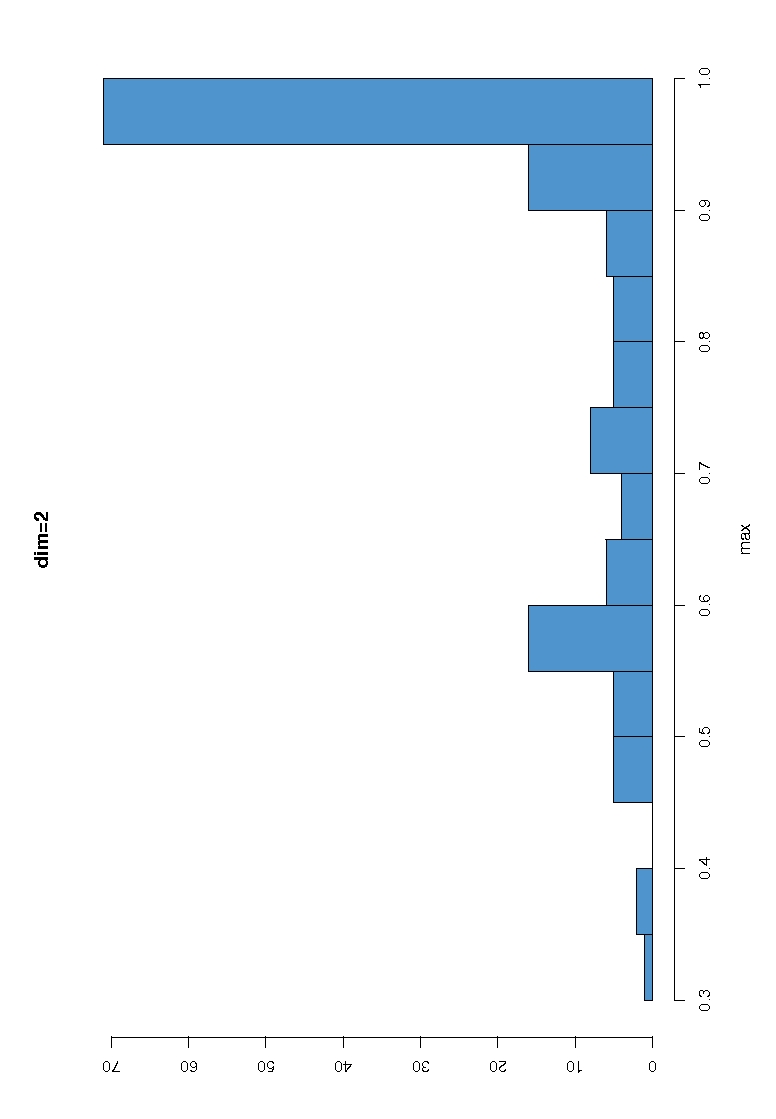}
\includegraphics[width=4cm,height=6cm,angle=270]{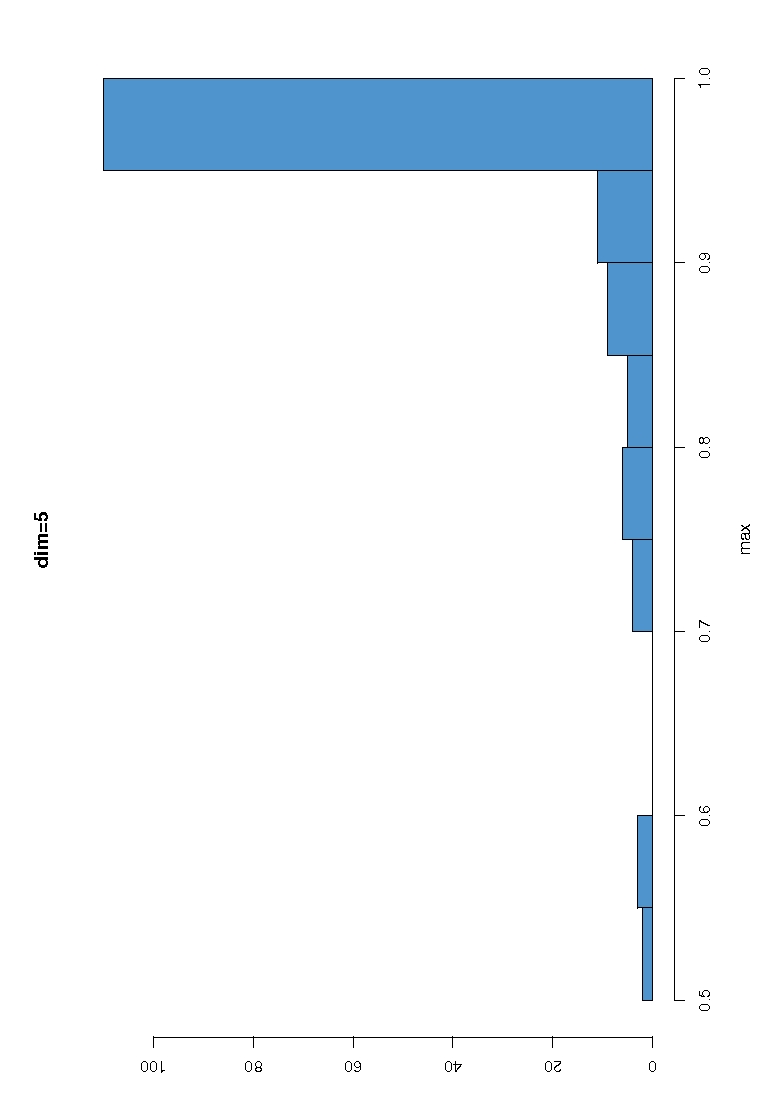}\hglue -.1cm
\includegraphics[width=4cm,height=6cm,angle=270]{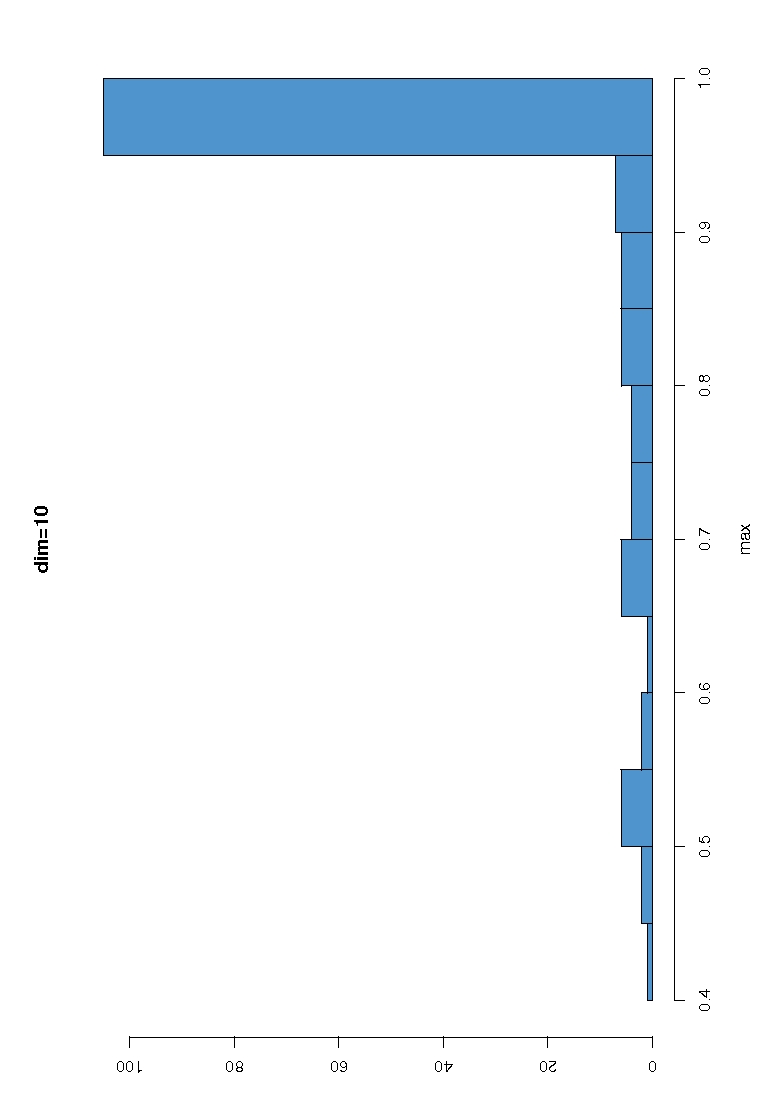}
\caption{\label{fig:dimcomp} Comparison of the distribution of the largest
importance weight based upon $150$ replications of an importance sampling
experiment with $245$ observations and dimensions $p=1,2,5,10$.}
\end{figure}
\end{exemple}

\subsection{Approximations of the Bayes factor}\index{Bayes factor!approximation|(}
As explained in Sections \ref{sub:bayesian:test} and \ref{sub:bayesian:mocho}, 
the first computational difficulty associated with Bayesian testing is 
the derivation of the {\em Bayes factor}, of the form
$$
B^\pi_{12} = \dfrac{ \displaystyle{ \int_{\Theta_1} f_1(x|\theta_1) \pi_1(\theta_1) \text{d}\theta_1 }
}{ \displaystyle{ \int_{\Theta_2} f_2(x|\theta_2) \pi_2(\theta_2) \text{d}\theta_2 } } =
\frac{m_1(x)}{m_2(x)}\,,
$$
where, for simplicity's sake, we have adopted a model choice perspective (that is,
$\theta_1$ and $\theta_2$ may live in completely different spaces).

Specific Monte Carlo methods for the estimation of ratios of normalising 
constants\index{normalising constant!ratios of}, or, equivalently, of Bayes factors,
have been developed in the past five years. See Chen {\em et al.} (2000, 
Chapter 5)\nocite{Chen:Shao:Ibrahim:2000} for a complete exposition, as well as
\cite{chopin:robert:2010} and \cite{marin:robert:2010} for recent reassessments.
In particular, the importance sampling technique is rather well-adapted 
to the computation of those\index{ratio!importance sampling for}
Bayes factors: Given a importance distribution, with density proportional 
to $g$, and a sample $\theta^{(1)},\ldots,\theta^{(T)}$ simulated from $g$, 
the marginal density for model $\mathfrak{M}_i$, $m_i(x)$, is approximated by
$$
\widehat{m}_i(x) = \displaystyle{ \sum_{t=1}^T  f_i(x|\theta^{(t)})
                    \dfrac{ \pi_i(\theta^{(t)}) }{ g(\theta^{(t)}) } }
                    \bigg/
  \displaystyle{ \sum_{t=1}^T \dfrac{ \pi_i(\theta^{(t)})}{
g(\theta^{(t)}) } } \,,
$$
where the denominator takes care of the (possibly) missing normalising 
constants\index{normalising constant}. (Notice that, if $g$ is a density, the expectation of
$\pi(\theta^{(t)})/g(\theta^{(t)})$ is $1$ and the denominator should be replaced by 
$T$ to decrease the variance of the estimator of $m_i(x)$.)

A compelling incentive, among others, for using importance sampling in the setting 
of model choice\index{importance sampling!for model choice} is that the sample 
	$(\theta^{(1)},\ldots,\theta^{(T)})$ 
can be recycled for all models $\mathfrak{M}_i$ sharing the same parameters (in
the sense that the models $\mathfrak{M}_i$ are parameterised in the same way, e.g.~by 
their first moments). 

\begin{exemple}[Continuation of Example \ref{ex:glm2}]\label{ex:glm3} 
In the case the $\beta$'s are simulated from a product instrumental distribution
$$
g(\beta) = \prod_{i=1}^p g_i(\beta_i)\,,
$$
the sample of $\beta$'s produced for the general model of Example \ref{ex:glm},
$\mathfrak{M}_1$ say, can also be used for the restricted model, $\mathfrak{M}_2$,
where $\beta_1=0$, simply by deleting the first component and
keeping the following components, with the corresponding importance density being
$$
g_{-1}(\beta) = \prod_{i=2}^p g_i(\beta_i)\,.
$$
Once the $\beta$'s have been simulated,\footnote{We stress the point that this is mostly an academic
exercise as, in regular settings, it is rarely the case that independent components are used for the
importance function.} the Bayes factor $B^\pi_{12}$ can be approximated by $\widehat{m}_1(x)/\widehat{m}_2(x)$.
\end{exemple}

However, the variance of $\widehat{m}(x)$ may be infinite, depending on the choice of $g$.
A possible choice is $g(\theta) = \pi(\theta)$, with wider tails than $\pi(\theta|x)$,
but this is often inefficient if the data is informative because the prior
and the posterior distributions will be quite different and most of the
simulated values $\theta^{(t)}$ fall outside the modal region of the
likelihood. (This is of course impossible when $\pi$ is improper.) For the choice $g(\theta) = f(x|\theta) \pi(\theta)$,
\begin{equation}\label{eq:IS.choiz}
  \widehat{m}(x) = 1 \bigg/\dfrac{1}{ T} \sum_{t=1}^T
                                 \dfrac{1 }{ f(x|\theta^{(t)})} \,,
\end{equation}
is the {\it harmonic mean}\index{harmonic mean} of the likelihoods, but
the corresponding variance is infinite when the likelihood has thinner 
tails than the prior (which is often the case). Having an infinite variance means that
the numerical value produced by the estimate cannot be trusted at all, as it may be away
from the true marginal by several orders of magnitude. As discussed in 
\cite{chopin:robert:2010} and \cite{marin:robert:2010}, it is actually possible to use
a generalised harmonic mean estimate
$$
  \widehat{m}(x) = 1 \bigg/\dfrac{1}{ T} \sum_{t=1}^T
                                 \dfrac{\varphi(\theta^{(t)})  }{ \pi(\theta^{(t)}) f(x|\theta^{(t)})} \,,
$$
when $\varphi$ is a probability density with tails that are thinner than the posterior tails. For instance,
a density with support an approximate HPD region is quite appropriate.

Explicitly oriented towards the computation of ratios of normalising
constants, {\em bridge sampling}\index{sampling!bridge} was introduced in 
\cite{Meng:Wong:1996}: if both models 
$\mathfrak{M}_1$ and $\mathfrak{M}_2$ cover the same parameter
space $\Theta$, if $\pi_1(\theta|x) = c_1 {\tilde\pi}_1(\theta|x)$
and $\pi_2(\theta|x) = c_2 {\tilde\pi}_2(\theta|x)$, where $c_1$ and $c_2$
are unknown normalising constants, then the equality
$$
   \frac{c_2}{c_1} = 
\frac{\mathbb{E}^{\pi_2}[{\tilde\pi}_1(\theta|x)\,h(\theta)]}
	{\mathbb{E}^{\pi_1}[{\tilde\pi}_2(\theta|x)\,h(\theta)]}
$$
holds for any {\it bridge function} $h(\theta)$ such that both expectations
are finite. The {\em bridge sampling} estimator is then
$$
B^S_{12} = \dfrac{ \displaystyle{ \dfrac{1}{ n_1} \sum_{i=1}^{n_1}
  {\tilde\pi}_2(\theta_{1i}|x) \,h(\theta_{1i}) }  }{
  \displaystyle{ \dfrac{1}{ n_2} \sum_{i=1}^{n_2} 
{\tilde\pi}_1(\theta_{2i}|x) \,h(\theta_{2i}) } }  \,,
$$
where the $\theta_{ji}$'s are simulated from $\pi_j(\theta|x)$
$(j=1,2,\,i=1,\ldots,n_j)$. 

For instance, if
$$
h(\theta) = 1/\left[ {\tilde\pi}_1(\theta|x) {\tilde\pi}_2(\theta_{1i}|x) \right]\,,
$$
then $B^S_{12}$ is a ratio of harmonic means\index{estimator!harmonic mean}, 
generalising (\ref{eq:IS.choiz}). \cite{Meng:Wong:1996}
have derived an (asymptotically) optimal bridge function\index{bridge!estimator}
$$
h^*(\theta) = \dfrac{n_1+n_2 }{ n_1 \pi_1(\theta|x) + n_2 \pi_2(\theta|x)}\,.
$$      
This choice is not of direct use, since the normalising constants of
$\pi_1(\theta|x)$ and $\pi_2(\theta|x)$ are unknown (otherwise, we should
not need to resort to such techniques!). Nonetheless, it shows that
a good bridge function should cover the support of both posteriors, with
equal weights if $n_1=n_2$.\index{bridge!sampling}

\begin{exemple}[Continuation of Example \ref{ex:glm}]\label{ex:glm4} 
For {\it generalised linear models}\index{model!generalised linear}, 
the mean (conditionally on the covariates) satisfies
$$
\mathbb{E}[y|\theta] = \nabla\psi(\theta) = \Psi(x^t\beta)\,,
$$
where $\Psi$ is the {\it link function}\index{function!link}.
The choice of the {\it link function}\index{function!link} $\Psi$ usually is 
quite open. For instance, when the $y$'s take values in $\{0,1\}$,
three common choices of $\Psi$ are \citep{McCullagh:Nelder:1989}
$$
\Psi_1(t) = \exp(t) / (1+\exp(t)),\quad
\Psi_2(t) = \Phi(t),\quad\hbox{and}\quad
\Psi_3(t) = 1 - \exp(-\exp(t))\,,
$$
corresponding to the {\em logit}, {\em probit} and {\it log--log} link functions
(where $\Phi$ denotes the c.d.f.\ of the $\mathscr{N}(0,1)$ distribution).
If the prior distribution $\pi$ on the $\beta$'s is a normal $\mathcal{N}_p(\xi,\tau^2 I_p)$,
and if the bridge function is $h(\beta)=1/\pi(\beta)$, the bridge 
sampling estimate is then $(1\le i<j\le 3)$
$$
B^S_{ij} = \dfrac{ \displaystyle{ \dfrac{1}{ n} \sum_{t=1}^{n} 
f_j(\beta_{it}|x) }  }{ \displaystyle{ \dfrac{1}{ n} \sum_{t=1}^{n} f_i(\beta_{jt}|x) } }\,,
$$
where the $\beta_{it}$ are generated from $\pi_i(\beta_i|x)\propto f_i(\beta_i|x)\pi(\beta_i)$, 
that is, from the true posteriors for each link function.\end{exemple}

The drawback in using bridge sampling is that the extension of the method to cases where 
the two models $\mathfrak{M}_1$ and $\mathfrak{M}_2$ do not cover the same parameter space,
for instance because the two corresponding parameter spaces are of different dimensions,
requires the introduction of a pseudo-posterior distribution \citep{Chen:Shao:Ibrahim:2000, marin:robert:2010} 
that complete the smallest parameter space so that dimensions match.\footnote{Section
\ref{sub:mcmc:reJ} covers in greater details the setting of varying dimension problems,
with the same theme that completion distributions and parameters are necessary but influential
for the performances of the approximation.} The impact of those completion pseudo-posteriors on
the quality of the Bayes factor approximation is such that they cannot be suggested for a general
usage in Bayes factor computation.

\begin{figure}
\centerline{\includegraphics[height=5cm]{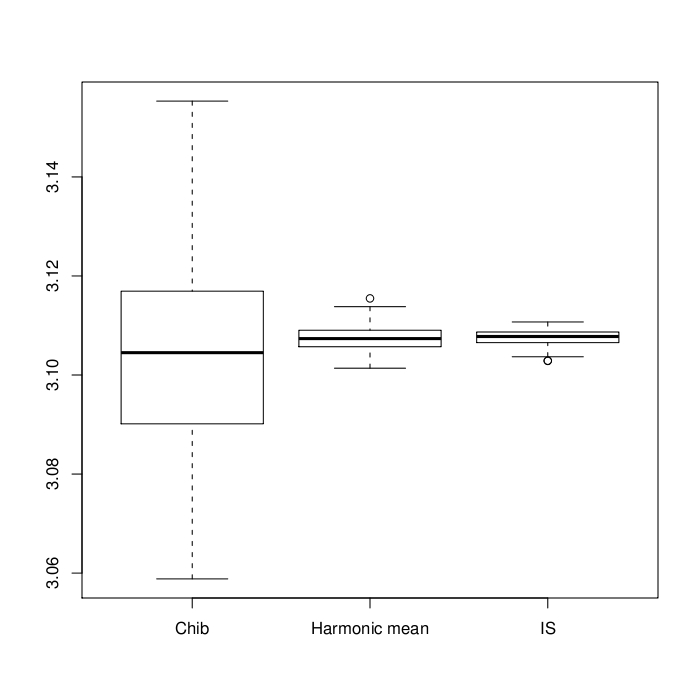}}
\caption{\label{fig:bfchi} In the setting of the probit modelling of {\sf R} Pima Indian dataset, 
using three covariates and testing for the significance of the diabetes pedigree function,\index{dataset!Pima Indian diabetes}
boxplots of 100 Chib's, harmonic mean and importance estimates of $B_{01}(y)$, based on 
simulations from the prior distributions, for $2\times 10^4$ simulations ({\em Source:} \citealp{marin:robert:2010}).}
\end{figure}

A completely different route to the approximation of a marginal likelihood is\index{likelihood!marginal}
Chib's (\citeyear{chib:1995}), which happens to be a direct application of
Bayes' theorem: given $x\sim f_k(x|\theta_k)$ and $\theta_k\sim\pi_k(\theta_k)$ $(k=1,2)$, we have that
$$
m_k(x) = \frac{f_k(x|\theta)\,\pi_k(\theta)}{\pi_k(\theta|x)}\,,
$$
for every value of $\theta$ (since both the lhs and the rhs of this equation are constant in $\theta$). Therefore, if an
arbitrary value of $\theta$, say $\theta^*_k$, is selected---e.g., the MAP estimate---and if a good approximation to $\pi_k(\theta|y)$
can be constructed, denoted $\hat{\pi}(\theta|y)$, Chib's (\citeyear{chib:1995})\index{Bayes factor!Chib's approximation}\index{Chib's 
approximation}\index{evidence} approximation to the marginal likelihood $m_k(x)$ is
\begin{equation}\label{eq:chib}
\widehat m_k(x) = \frac{f_k(y|\theta^*_k)\,\pi_k(\theta^*_k)}{\hat{\pi_k}(\theta^*_k|y)}\,.
\end{equation}
As a first solution, $\hat{\pi}(\theta|y)$ may be a Gaussian approximation based on the MLE,
but this is unlikely to be accurate in a general setting. Chib's (\citeyear{chib:1995}) solution 
is to use a nonparametric approximation based on a preliminary Monte Carlo Markov chain (Section \ref{sec:mcmc}) 
sample, even though the accuracy may\index{variable!latent}
also suffer in large dimensions.  In the special setting of latent variables models (like the mixtures of 
distributions discussed in Example \ref{ex:mix}), this approximation is particularly attractive as there 
exists a natural approximation to $\pi_k(\theta|y)$, based on the Rao--Blackwell
\citep{Gelfand:Smith:1990} estimate
$$
\hat{\pi_k}(\theta^*_k|y) = \frac{1}{T}\,\sum_{t=1}^T \pi_k(\theta^*_k|y,z_k^{(t)})\,,
$$
where the $z_k^{(t)}$'s are the latent variables simulated by the Monte Carlo Markov chain algorithm.
The estimate $\hat{\pi_k}(\theta^*_k|y)$
is a parametric unbiased approximation of $\pi_k(\theta^*_k|y)$ that converges to the true density with rate $\text{O}(\sqrt{T})$.
This Rao--Blackwell approximation obviously requires the full conditional density $\pi_k(\theta^*_k|y,z)$ to be
available in closed form (constant included) but this is the case for many standard models. Figure \ref{fig:bfchi} reproduces
an illustration from \cite{marin:robert:2010} that compares the variability of Chib's (1995) method against nearly optimal harmonic
and importance sampling approximations in the setting of a probit posterior distribution. While the former solution varies 
more than the two latter solutions, its generic features make Chib's (1995) method a reference for this problem.

While amenable to an importance sampling technique of sorts,
the alternative approach of nested sampling \citep{skilling:2007a} for computing evidence
is discussed in \cite{chopin:robert:2010} and \cite{robert:wraith:2009}. Similarly, the
specific approach based on the Savage-Dickey \citep{dickey:1971,verdinelli:wasserman:1995} 
representation of the Bayes factor associated with the null hypothesis $H_0:\,\theta=\theta_0$, as
$$
B_{01}(x) = \dfrac{\pi_1(\theta_0|x)}{\pi_1(\theta_0)}\,,
$$
can be related to the family of bridge sampling techniques, even though the initial theoretical foundations of the
method are limited, as explained in \cite{robert:marin:2009}. This paper engineered a general framework that 
produces a converging and unbiased approximation of the Bayes factor, unrelated
with the approach of \cite{verdinelli:wasserman:1995}, that builds upon a mixed pseudo-posterior naturally
derived from the priors under both the null and the alternative hypotheses.

As can be seen from the previous developments, such methods require a rather careful tuning
to be of any use. Therefore, they are rather difficult to employ outside settings where pairs
of models are opposed. In other words, they cannot be directly used in general model choice
settings where the parameter space (and in particular the parameter dimension) varies across
models, as in, for instance, Example \ref{ex:AR2}. 
To address the computational issues corresponding to these cases requires more advanced
techniques introduced in the next Section.\index{Bayes factor!approximation|)}\index{Monte Carlo techniques|)}

\section{Markov Chain Monte Carlo Methods}\label{sec:mcmc}

As described precisely in Chapter 
II.3 and in \cite{robert:casella:2004}, MCMC methods try to overcome the limitation of
regular Monte Carlo methods through the use of a Markov chain with stationary
distribution the posterior distribution. There exist rather generic
ways of producing such chains, including Metropolis--Hastings and Gibbs
algorithms. Besides the fact that stationarity of the target\index{distribution!target}
distribution is enough to justify a simulation method by Markov chain generation,
the idea at the core of MCMC algorithms is that local exploration, when properly
weighted, can lead to a valid representation of the distribution of interest, as
for instance, when using the Metropolis--Hastings algorithm.\index{algorithm!Metropolis--Hastings|(}
\index{Markov chain Monte Carlo (MCMC) algorithm|(}

\subsection{Metropolis--Hastings as universal simulator}\label{sub:mcmc:uni}
The Metropolis--Hastings, presented in \cite{robert:casella:2004} and Chapter 
II.3, offers a straightforward solution to the problem of simulating from the posterior
distribution $\pi(\theta|x)\propto f(x|\theta)\,\pi(\theta)$: starting from
an arbitrary point $\theta_0$, the corresponding Markov chain explores the
surface of this posterior distribution by a similarly arbitrary random walk proposal $q(\theta|\theta^\prime)$
that progressively visits the whole range of the possible values of $\theta$.

\bigskip 
\begin{minipage}[hbt]{\textwidth}
\centerline{\bf ---Metropolis--Hastings Algorithm---}
\smallskip 
\colorbox{LightGrey}{\makebox[0.9\textwidth][l]{\parbox{0.85\textwidth}{
{\sf At iteration $t$
\begin{enumerate}
\item Generate $\xi\sim q( \xi | \theta^{(t)} )$, $u_t\sim\mathcal{U}([0,1])$
\item Take
$$
\theta^{(t+1}) = \begin{cases} \xi_t &\hbox{if } u_t\le 
\displaystyle{
\frac{\pi(\xi_t|x)}{\pi(\theta^{(t)}|x)} 
\frac{q(\theta^{(t)}|\xi_t)}{q(\xi_t|\theta^{(t)})}}\cr
	\theta^{(t)} &\hbox{otherwise} \end{cases}
$$
\end{enumerate}
}}}}
\end{minipage}\hfill

\begin{figure}[hbtp]
\includegraphics[height=10cm,width=6cm,angle=270]{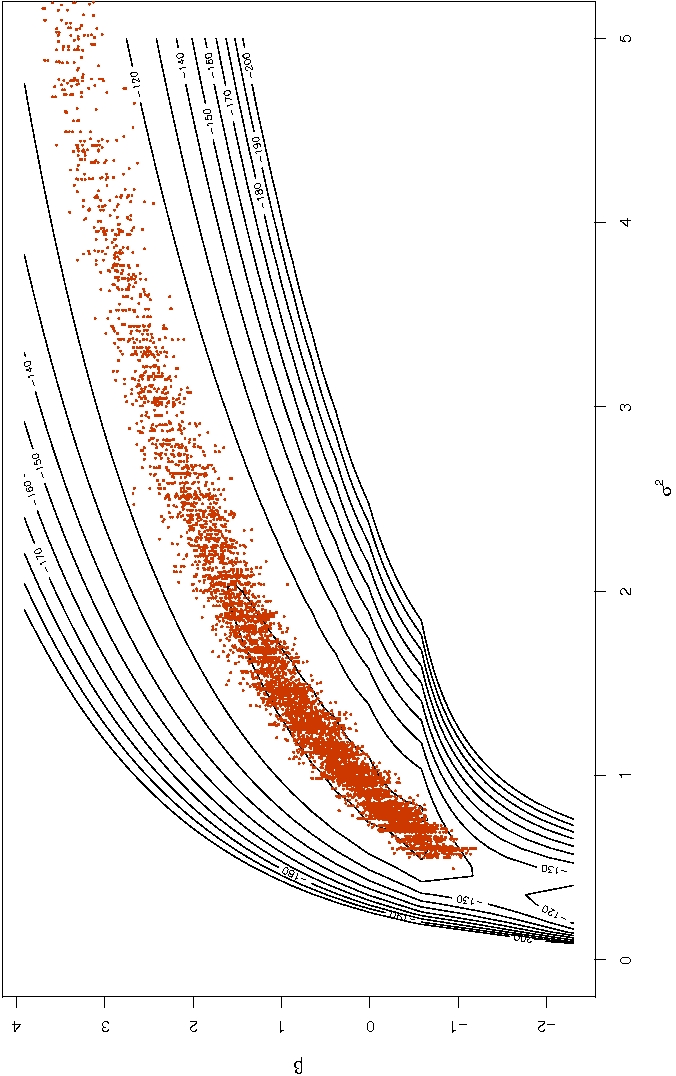}
\vglue -0.2truecm 
\caption[Probit log-posterior]{\label{fig:newpro1}
Contour plot of the log-posterior distribution for a probit sample of $1,000$
observations, along with $1,000$ points of an MCMC sample {\em (Source:
\citealp{robert:casella:2004})}.}
\end{figure}

\begin{figure}\begin{center} 
\includegraphics[height=6cm,width=8cm]{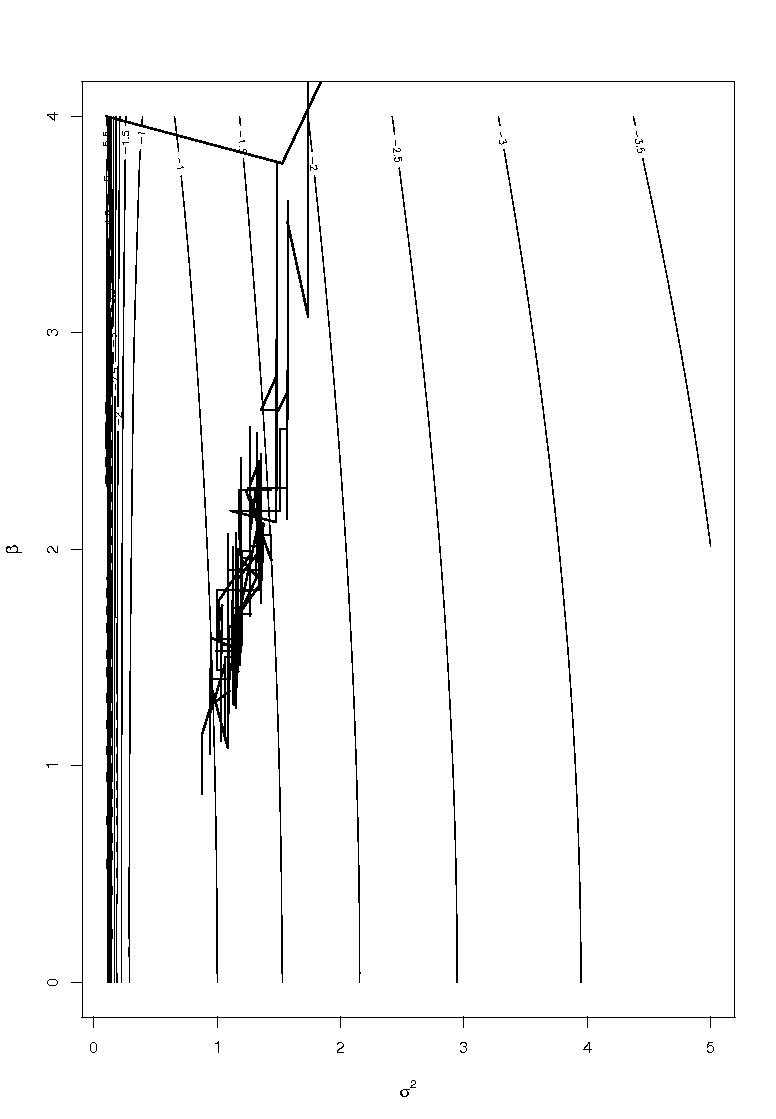}
\end{center}
\vglue -0.2truecm 
\caption[First Metropolis-Hastings steps on the probit log-posterior]{\label{fig:newpro2}
First $500$ steps of the Metropolis--Hastings algorithm on the probit log-posterior
surface, when started at $(\beta,\sigma^2)=(0.1,4.0)$.}
\end{figure}

\medskip
\begin{exemple}[Continuation of Example \ref{ex:glmPbit}]\label{ex:glmPbit2}
In the case $p=1$, the probit model defined in Example \ref{ex:glmPbit} can also 
be over-parameterised as\index{model!probit} 
$$
\text{Pr}(Y_i=1|x_i) = 1 - \text{Pr}(Y_i=0|x_i) = \Phi(x_i\beta/\sigma)\,,
$$
since it only depends on $\beta/\sigma$. The Bayesian processing of non-identified
models poses no serious difficulty as long as the posterior distribution is well
defined. This is the case for a proper prior like
$$
\pi(\beta,\sigma^2) \propto \sigma^{-4}\,\exp\{-1/\sigma^2\}\,\exp\{-\beta^2/50)
$$
that corresponds to a normal distribution on $\beta$ and a gamma distribution
on \smash{$\sigma^{-2}$}.
While the posterior distribution on $(\beta,\sigma)$ is not a standard distribution,
it is available up to a normalising constant. Therefore, it can be directly processed via
an MCMC algorithm. In this case, we chose a Gibbs sampler that simulates $\beta$ and $\sigma^2$ 
alternatively, from
$$
\pi(\beta|\mathbf{x},\mathbf{y},\sigma) \propto \prod_{y_i=1} \Phi(x_i\beta/\sigma)
\prod_{y_i=0} \Phi(-x_i\beta/\sigma) \times\pi(\beta)
$$
and
$$
\pi(\sigma^2|\mathbf{x},\mathbf{y},\beta) \propto \prod_{y_i=1} \Phi(x_i\beta/\sigma)
\prod_{y_i=0} \Phi(-x_i\beta/\sigma) \times\pi(\sigma^2)
$$
respectively.  Since both of these conditional distributions are also non-standard, we 
replace the direct simulation by a one-dimensional Metropolis--Hastings step, using
normal $\mathcal{N}(\beta^{(t)},1)$
and log-normal $\mathcal{L}\mathcal{N}(\log\sigma^{(t)},.04)$
random walk proposals, respectively. For a simulated dataset of $1,000$ points,
the contour plot of the log-posterior distribution is given in Figure \ref{fig:newpro1},
along with the last $1,000$ points of a corresponding MCMC sample after $100,000$ iterations.
This graph shows a very satisfactory repartition of the simulated parameters over the likelihood
surface, with higher concentrations near the largest posterior regions. For another simulation,
Figure \ref{fig:newpro2} details the first $500$ steps, when started
at $(\beta,\sigma^2)=(0.1,4.0)$. Although each step contains both a $\beta$ {\em and} a $\sigma$
proposal, some moves are either horizontal or vertical: this corresponds to cases when either
the $\beta$ or the $\sigma$ proposals have been rejected. Note also the fairly rapid convergence
to a modal zone of the posterior distribution in this case.\end{exemple}


Obviously, this is only a toy example and more realistic probit models do not
fare so well with down-the-shelf random walk Metropolis--Hastings algorithms,
as shown for instance in \cite{Nobile:1998} (see also \citealp{robert:casella:2004}, 
Section 10.3.2, \citealp{marin:robert:2007}, Section 4.3, and \citealp{marin:robert:2010}).\footnote{Even 
in the simple case of the probit model, MCMC algorithms do 
not always converge very quickly, as shown in \citeauthor{robert:casella:2004} (2004, Chapter 14).}

The difficulty inherent to random walk Metropolis--Hastings algorithms is
the scaling\index{algorithm!scaling}
of the proposal distribution: it must be adapted to the shape
of the target distribution so that, in a reasonable number of steps,  the
whole support of this distribution can be visited. If the scale of the
proposal is too small, this will not happen as the algorithm stays ``too
local" and, if there are several modes on the posterior, the algorithm\index{mode!multiple}
may get trapped within one modal region because it cannot reach other
modal regions with jumps of too small a magnitude. The larger the dimension $p$
is, the harder it is to set up the right scale, though, because 
\begin{enumerate}\renewcommand{\theenumi}{(\alph{enumi})}
\item the curse of dimension\index{curse of dimension}
implies that there are more and more empty 
regions in the space, that is, regions with zero posterior probability; 
\item the knowledge and intuition about the modal regions get weaker and weaker;
\item the proper scaling involves a symmetric $(p,p)$ matrix $\Xi$ in the
proposal $g((\theta-\theta^\prime)^\Tee \Xi (\theta-\theta^\prime))$.
Even when the matrix $\Xi$ is diagonal, it gets harder to scale as the
dimension increases (unless one resorts to a Gibbs like implementation,
where each direction is scaled separately). 
\end{enumerate} Note also that the on-line\index{scaling!on-line}\index{scaling!adaptive} 
scaling of the algorithm against the empirical acceptance rate is\index{acceptance rate!empirical}
inherently flawed in that (a) the process is no longer Markovian and (b)
the attraction of a modal region may give a
false sense of convergence and lead to a choice of too small a scale,\index{mode!attraction}
simply because other modes will not be visited during the scaling
experiment.\index{algorithm!Metropolis--Hastings|)}\index{acceptance rate!empirical}  

\subsection{Gibbs sampling and latent variable models}\label{sub:mcmc:late}
The Gibbs sampler\index{Gibbs sampler|(}\index{Bayesian!hierarchical structures} 
is a definitely attractive algorithm for Bayesian problems\index{variable!latent|(}
because it naturally fits the hierarchical structures so often found in such
problems. ``Natural" being a rather vague notion from a simulation point of 
view, it routinely happens that other algorithms fare better than the Gibbs
sampler. Nonetheless, Gibbs sampler is often worth a try (possibly with other
Metropolis--Hastings refinements at a later stage) in well-structured objects
like Bayesian hierarchical models and more general graphical models. 

A very relevant illustration is made of latent variable models, where the observational
model is itself defined as a mixture model,\index{distribution!mixture}\index{mixture!and latent variables}
$$
f(x|\theta) = \int_\mathscr{Z} f(x|z,\theta)\,g(z|\theta)\,\text{d} z.
$$
Such models were instrumental in promoting the Gibbs sampler in the sense that they
have the potential to make Gibbs sampling sound natural very easily. (See also Chapter 
II.3.)
For instance, \cite{Tanner:Wong:1987}
wrote a precursor article to \cite{Gelfand:Smith:1990} that designed specific two-stage
Gibbs samplers for a variety of latent variable models. And many of the first applications
of Gibbs sampling in the early 90's were actually for models of that kind.
The usual implementation of the Gibbs sampler in this
case is to simulate the missing variables $Z$ conditional on the parameters and
reciprocally, as follows:\index{missing variables!simulation of}

\bigskip
\centerline{\bf ---Latent Variable Gibbs Algorithm---}\nopagebreak[4]
\smallskip
\colorbox{LightGrey}{\makebox[0.9\textwidth][l]{\parbox{0.85\textwidth}{
\medskip
{\sf
At iteration $t$
\begin{enumerate}
\item Generate $z^{(t+1)} \sim g(z|\theta^{(t)},x)$
\item Generate $\theta^{(t+1)} \sim \pi(\theta|x,z^{(t+1)})$
\end{enumerate}
}
}}}

\bigskip
While we could have used the probit case as an illustration (Example \ref{ex:glmPbit}),
as done in Chapter 
II.3,
we choose to pick the case of mixtures (Example \ref{ex:mix}) as a better setting.

\begin{exemple}[Continuation of Example \ref{ex:mix}]\label{ex:mix2}
\index{mixtures!and latent variables}The natural missing data structure of a mixture of distribution is
historical. In one of the first mixtures to be ever studied by Bertillon,
in 1863,\index{Bertillon, 
Alphonse} a bimodal structure on the height of conscripts in south eastern France (Doubs)
can be explained by the mixing of two populations of military conscripts, 
one from the plains and one from the mountains (or hills). Therefore, in the analysis 
of data from distributions of the form\index{military conscripts}
$$
\sum_{i=1}^k p_i f(x|\theta_i)\,,
$$
a common missing data representation is to associate with each observation $x_j$ a
missing multinomial variable $z_j\sim \mathcal{M}_k(1;p_1,\ldots,p_k)$ such that
$x_j|z_j=i\sim f(x|\theta_i)$. In heterogeneous populations made of several
homogeneous subgroups or subpopulations, it makes sense to interpret $z_j$ as
the index of the population of origin of $x_j$, which has been lost in the
observational process.\index{heterogeneous populations}

However, mixtures are also customarily used for density approximations, as a limited
dimension proxy to non-parametric approaches. In such cases, the components of the
mixture and even the number $k$ of components in the mixture are often meaningless
for the problem to be analysed. But this distinction between natural and artificial
completion (by the $z_j$'s) is lost to the MCMC sampler, whose goal is simply to
provide a Markov chain that converges to the posterior as stationary distribution.
Completion is thus, from a simulation point of view, a mean to generate such a 
chain.\index{completion}\index{model!mixture}

The most standard Gibbs sampler for mixture models \citep{Diebolt:Robert:1994} is
thus based on the successive simulation of the $z_j$'s and of the $\theta_i$'s,
conditional on one another and on the data:
\begin{enumerate}
\renewcommand\theenumi{\arabic{enumi}.}
\item Generate $z_j|\theta,x_j$ $(j=1,\ldots,n)$
\item Generate $\theta_i|\mathbf{x},\mathbf{z}$ $(i=1,\ldots,k)$
\end{enumerate}
Given that the density $f$ is most often from an exponential family, the simulation
of the $\theta_i$'s is generally straightforward.

As an illustration, consider the (academic) case of a normal mixture with two components,
with equal known variance and fixed weights,
\begin{equation}\label{eq:gimi9}
p\,\mathcal{N}(\mu_1,\sigma^2) + (1-p)\,\mathcal{N}(\mu_2,\sigma^2) \,.
\end{equation}
Assume in addition a normal $\mathcal{N}(0,10\sigma^2)$ prior on both
means $\mu_1$ and $\mu_2$. It is easy to see that
$\mu_1$ and $\mu_2$ are independent, given $(\mathbf{z},\mathbf{x})$, 
and the respective conditional distributions are
$$
\mathcal{N} \left( \sum_{z_i=j} x_i / \left( .1 + n_j\right), 
	\sigma^2 / \left( .1 + n_j\right) \right)\,,
$$
where $n_j$ denotes the number of $z_i$'s equal to $j$. Even more easily, 
it comes that the conditional posterior
distribution of $\mathbf{z}$ given $(\mu_1,\mu_2)$ is a product of binomials, with
\begin{eqnarray*}
\lefteqn{\text{Pr}(Z_i=1|x_i,\mu_1,\mu_2)} \\
&=& \frac{p \exp\{-(x_i-\mu_1)^2/2\sigma^2 \} }{ p \exp\{-(x_i-\mu_1)^2/2\sigma^2 \} + (1-p) \exp\{-(x_i-\mu_2)^2/2\sigma^2 \} }\,.
\end{eqnarray*}
Figure \ref{fig:gib+mix1} illustrates the behaviour of the Gibbs sampler in that
setting, with a simulated dataset of $100$ points from the $.7 \mathcal{N}(0,1)
+.3 \mathcal{N}(2.7,1)$ distribution. The representation of the MCMC sample after
$5,000$ iterations is quite in agreement with the posterior surface, represented via
a grid on the $(\mu_1,\mu_2)$ space and some contours. The sequence of consecutive 
steps represented on the left graph also shows that the mixing\index{Gibbs sampler!mixing of} 
behaviour is satisfactory, since the jumps are scaled in terms of the modal region of the posterior.
\begin{figure}[hbt]
\includegraphics[width=5cm,height=5cm]{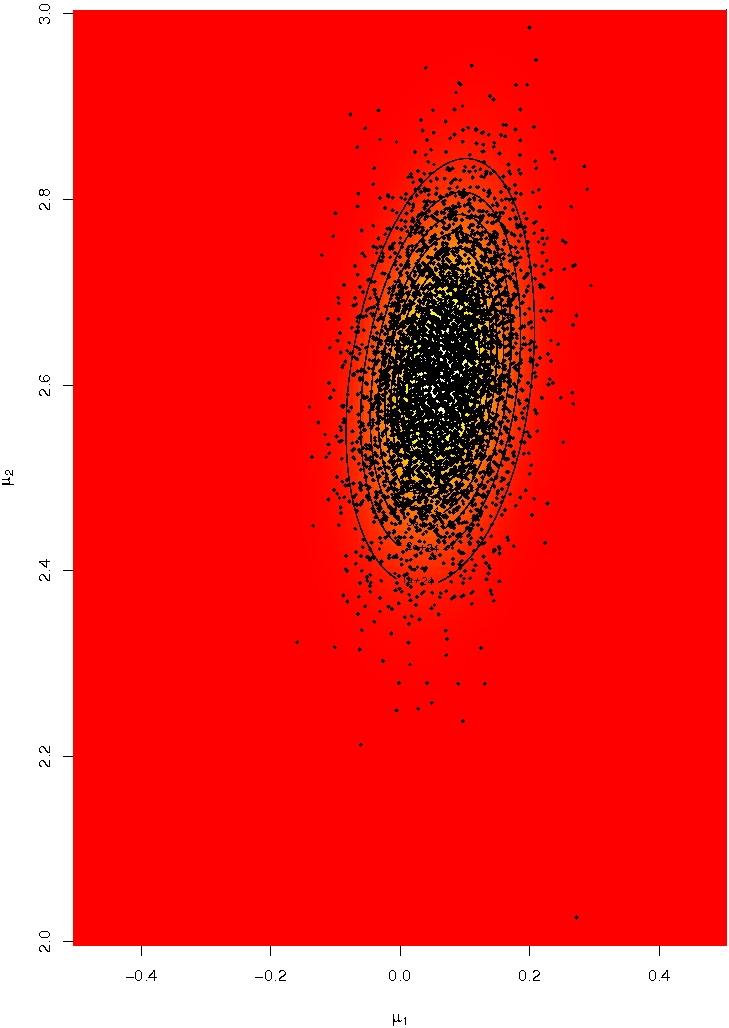}\qquad
\includegraphics[width=5cm,height=5cm]{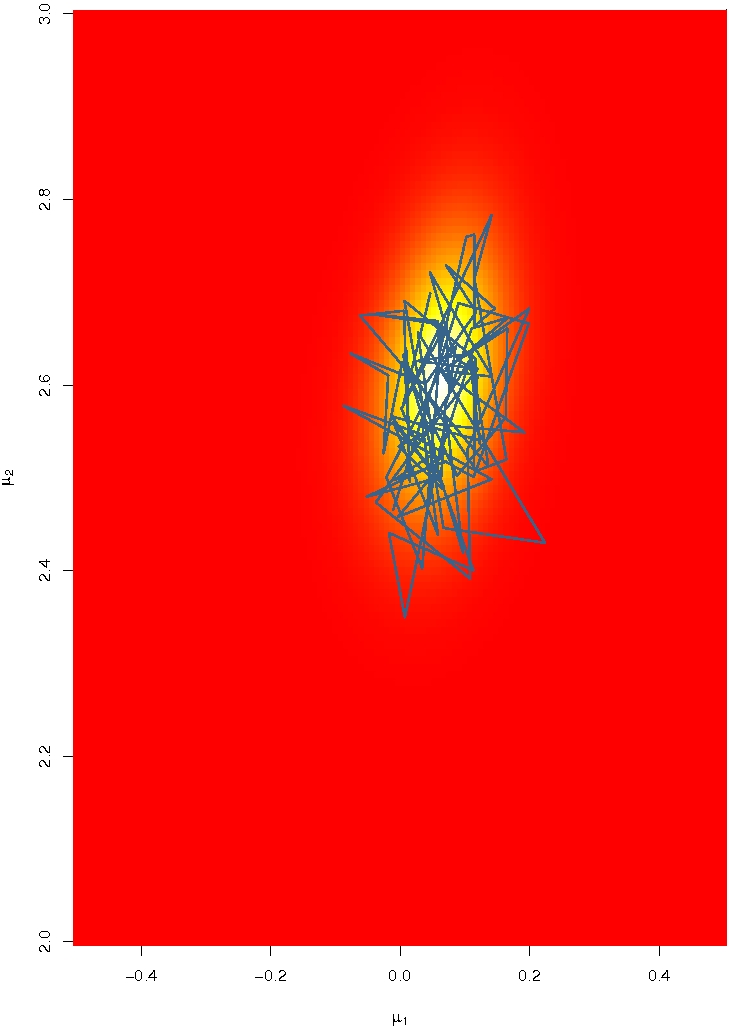}
\caption[Gibbs output for mixture posterior]{\label{fig:gib+mix1}
Gibbs sample of $5,000$ points for the mixture posterior {\em (left)} and
path of the last $100$ consecutive steps {\em (right)} against the posterior
surface {\em (Source: \citealp{robert:casella:2004})}.}
\end{figure}

This experiment gives a wrong sense of safety, though, because it does not point
out the fairly large dependence of the Gibbs sampler to the initial conditions,
already signalled in \cite{Diebolt:Robert:1994} under the name of {\em trapping
states}.\index{trapping state}
Indeed, the conditioning of $(\mu_1,\mu_2)$ on $\mathbf{z}$ implies that
the new simulations of the means will remain very close to the previous values,
especially if there are many observations, and thus that the new allocations $\mathbf{z}$
will not differ much from the previous allocations. In other words, to see a significant
modification of the allocations (and thus of the means) would require a very very large
number of iterations. Figure \ref{fig:wronmix} illustrates this phenomenon for the same
sample as in Figure \ref{fig:gib+mix1}, for a wider scale: there always exists a second mode in
the posterior distribution, which is much lower than the first mode located around $(0,2.7)$.
Nonetheless, a Gibbs sampler initialised close to the second and lower mode will not be
able to leave the vicinity of this (irrelevant) mode, even after a large number of
iterations. The reason is as given above: to jump to the other mode, a majority of $z_j$'s
would need to change simultaneously and the probability of such a jump is too close to
$0$ to let the event occur.\footnote{It is quite interesting to see
that the mixture Gibbs sampler suffers from the same pathology as the EM 
algorithm, although this is not surprising given that it is based 
on the same completion scheme.}\index{algorithm!EM}
\begin{figure}[hbtp] 
\includegraphics[height=11cm,width=5cm,angle=270]{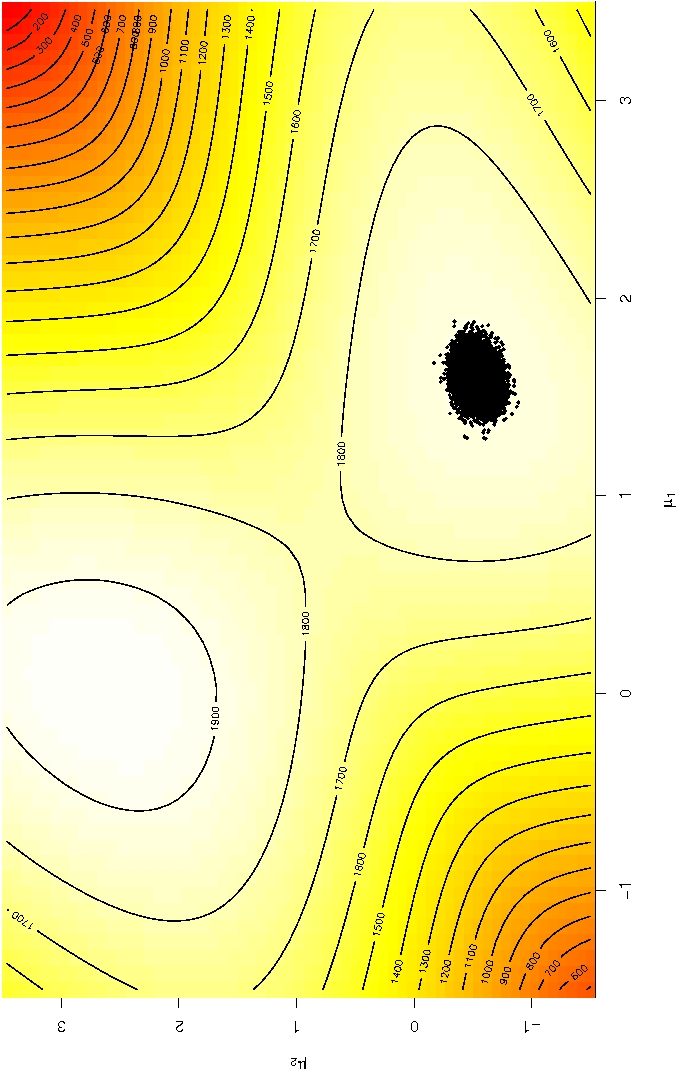}
\vskip -.3truecm
\caption[Gibbs sampler stuck at a wrong mode]{\label{fig:wronmix}
Posterior surface and corresponding Gibbs sample 
for the two mean mixture model, when initialised close to the second 
and lower mode, based on $10,000$ iterations {\em (Source: \citealp{robert:casella:2004})}.}
\end{figure}
The literature on this specific issue---of exploring all the modes of the posterior
distribution---has grown around the denomination of ``label switching problem"\index{label
switching problem}. For instance, \cite{celeux:hurn:robert:2000} have pointed out the
deficiency of most MCMC samplers in this respect, while 
\cite{fruhwirth:2001,fruhwirth:2004,fruhwirth:2006}, \cite{berkhof:mechelen:gelman:2003},
\cite{jasra:holmes:stephens:2005}, and \cite{geweke:2007} have proposed different devices
to overcome this difficulty, in particular when testing for the number of components. See
\cite{lee:marin:mengersen:robert:2008} for a survey on recent developments.
\end{exemple}

This example illustrates quite convincingly that, while the completion is natural
from a model point of view (since it is a part of the definition of the model), 
it does not necessarily transfer its utility for the simulation of the posterior.
Actually, when the missing variable model allows for a closed form likelihood, as 
is the case for mixtures, probit models (Examples \ref{ex:glmPbit} and 
\ref{ex:glmPbit2}) and even hidden Markov models \citep[see][]{Cappe:Ryden:2004},
the whole range of the MCMC technology can be used as well. The appeal of alternatives
like random walk Metropolis--Hastings schemes is that they remain in a smaller dimension
space, since they avoid the completion step(s), and that they are not restricted in
the range of their moves.\footnote{This wealth of possible alternatives to the completion
Gibbs sampler is a mixed blessing in that their range, for instance
the scale of the random walk proposals, needs to be scaled properly to avoid inefficiencies.}

\begin{exemple}[Continuation of Example \ref{ex:mix2}]\label{ex:mix3}
Given that the likelihood of a sample $(x_1,\ldots,x_n)$ from the mixture
distribution (\ref{eq:gimi9}) can be computed in $\hbox{O}(2n)$ time, a
regular random walk Metropolis--Hastings algorithm can be used in this setup. 
Figure \ref{fig:rwmix} shows how quickly this algorithm escapes the attraction of the
poor mode, as opposed to the Gibbs sampler of Figure \ref{fig:wronmix}: within a
few iterations of the algorithm, the chain drifts over the poor mode
and converges almost deterministically to the proper region of the
posterior surface. The random walk is based on $\mathcal{N}(\mu_i^{(t)},0.04)$
proposals, although other scales would work as well but would require more iterations
to reach the proper model regions. For instance, a scale of $0.005$ in the Normal
proposal above needs close to $5,000$ iterations to attain the main mode.
\begin{figure}[hbtp]
\includegraphics[height=11cm,width=7cm,angle=270]{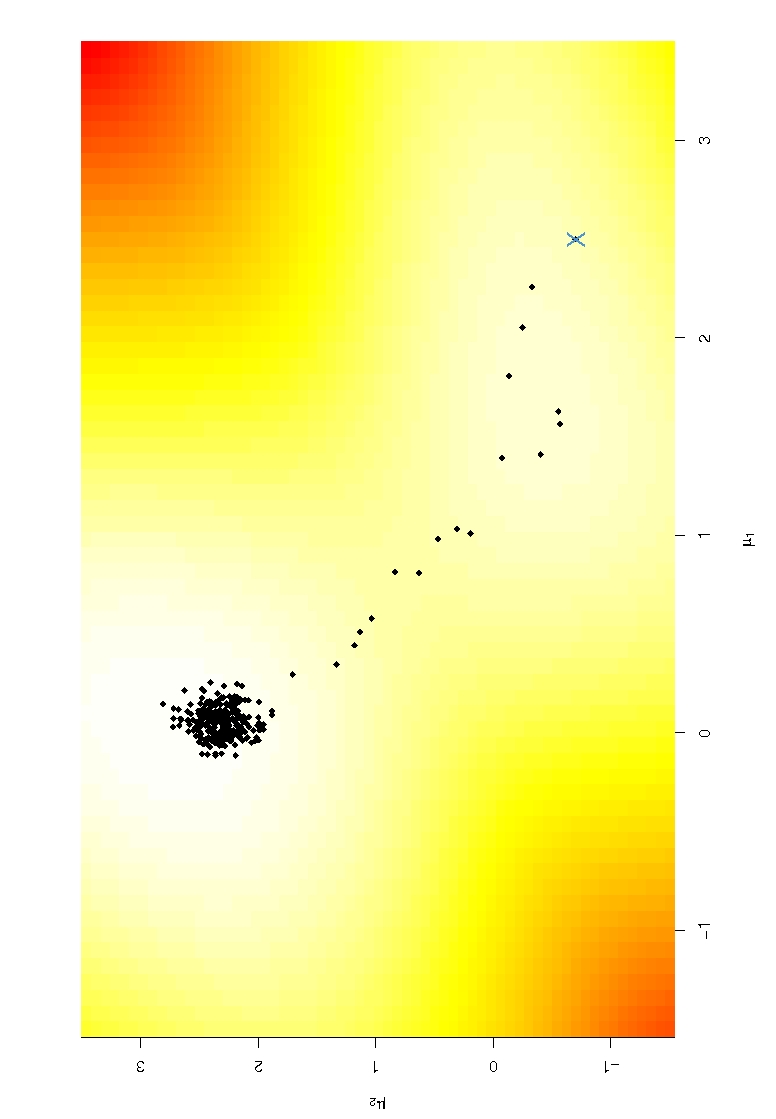}
\caption[RWHM sampler escaping a wrong mode]{\label{fig:rwmix}
Track of a $1,000$ iteration random walk Metropolis--Hastings chain
on the posterior surface, the starting point being indicated by a cross.
{\em (The scale of the random walk is $0.2$.)}}
\end{figure}
\end{exemple}

The secret of a successful MCMC implementation in such latent variable models is
to maintain the distinction between latency in models and latency in simulation
(the later being often called use of {\em auxiliary 
variables}).\index{variable!auxiliary} When latent
variables can be used with adequate mixing of the resulting chain and when 
the likelihood cannot be computed in a closed form (as in hidden semi-Markov
models, \citealp{Cappe:Guillin:Marin:Robert:2003}), a Gibbs sampler is a rather
simple solution that is often easy to simulate from. Adding well-mixing random 
walk Metropolis--Hastings steps in the simulation scheme
cannot hurt the overall mixing of the chain \citep[][Chap. 13]{robert:casella:2004},
especially when several scales can be used at once (see Section \ref{sec:MoMa}).
Note also that the technique of tempering\index{tempering} that flattens the target
distribution in order to facilitate its exploration by a Markov chain is available for
most latent variable models, if sometimes in rudimentary versions. See \cite{chopin:2007} and
Marin and Robert (2007, Section 6.6) for an illustration in the setups of hidden Markov models
and of mixtures (Example \ref{ex:mix3}), respectively.\index{hidden Markov models}
A final word is that the latent variable completion can be conducted in an infinity of ways and that
several of these should be tried or used in parallel to increase the chances
of success.\index{Gibbs sampler|)}\index{variable!latent|)}

\subsection{Reversible jump algorithms for variable dimension models}\label{sub:mcmc:reJ}
As described in Section \ref{sub:bayesian:mocho}, model choice is computationally
different from testing in that it considers at once a (much) wider range of models
$\mathfrak{M}_i$ and parameter spaces $\Theta_i$.\index{model!choice}
Although early approaches could only go through a pedestrian pairwise
comparison, a more adequate perspective is to envision the model index $\mu$
as part of the parameter to be estimated, as in (\ref{eq:bigset}).\index{dimension!unknown}
The (computational) difficulty is that we are then dealing with a possibly
infinite space\footnote{The difficulty with the infinite part of the problem is easily solved in that the setting
is identical to simulation problems in (countable or uncountable) infinite spaces. When
running simulations in those spaces, some values are never visited by the simulated 
Markov chain and the chances a value is visited is related with the probability of this
value under the target distribution.}\index{Markov chain Monte Carlo (MCMC) algorithm!reversible}
that is the collection of unrelated sets: how can we then
simulate from the corresponding distribution?\footnote{Early proposals to solve
the varying dimension problem involved saturation schemes where all the parameters for
all models were updated deterministically \citep{Carlin:Chib:1995}, but
they do not apply for an infinite collection of models and they need to\index{model!index}
be precisely calibrated to achieve a sufficient amount of moves between models.}

The MCMC solution proposed\index{algorithm!Green's}\index{reversible jump MCMC|(}
by \cite{Green:1995} is called {\em reversible jump MCMC}, because
it is based on a {\it reversibility} constraint on the transitions
between the sets $\Theta_i$. In fact, the only real difficulty compared
with previous developments is to validate moves (or {\em jumps\/})
between the $\Theta_i$'s, since proposals restricted to a given 
$\Theta_i$ follow from the usual (fixed-dimensional) theory.
Furthermore, {\it reversibility} can be processed at a local level:
since the model indicator $\mu$ is a integer-valued random variable, 
we can impose reversibility for each pair $(k_1,k_2)$ of possible
values of $\mu$. The idea at the core of reversible jump MCMC is then
to supplement each of the spaces $\Theta_{k_1}$ and
$\Theta_{k_2}$ with adequate artificial spaces in order to create a 
{\em bijection}\index{function!one-to-one} between them.
For instance, if $\dim(\Theta_{k_1})>\dim(\Theta_{k_2})$ and if the
move from ${\Theta}_{k_1}$ to ${\Theta}_{k_2}$ can be represented
by a {\em deterministic} transformation of $\theta^{(k_1)}$
$$
\theta^{(k_2)}=T_{k_1\rightarrow k_2}(\theta^{(k_1)})\,,
$$
Green (1995) imposes a {\it dimension matching}\index{dimension!matching}
condition which is that the opposite move from ${\Theta}_{k_2}$ 
to ${\Theta}_{k_1}$ is concentrated on the curve
$$
\left\{ \theta^{(k_1)}\,:\,\theta^{(k_2)}
=T_{k_1\rightarrow k_2}(\theta^{(k_1)}) \right\} \;.
$$
In the general case, if $\theta^{(k_1)}$ is completed by a
simulation $u_{k_1}\sim g_{k_1}(u_{k_1})$ into $(\theta^{(k_1)},\allowbreak u_{k_1})$
and $\theta^{(k_2)}$ by $u_{k_2}\sim g_{k_2}(u_{k_2})$ into $(\theta^{(k_2)},\allowbreak u_{k_2})$ 
so that the mapping
between $(\theta^{(k_1)},u_{k_1})$ and $(\theta^{(k_2)},u_{k_2})$ is a bijection, 
\begin{equation}\label{eq:ze_transf}
(\theta^{(k_2)},u_{k_2}) = T_{k_1\rightarrow k_2}(\theta^{(k_1)},u_{k_1}),
\end{equation}
the probability of acceptance for the move from model ${\mathfrak M}_{k_1}$
to model ${\mathfrak M}_{k_2}$ is then
$$
\min\left( \dfrac{\pi(k_2,\theta^{(k_2)}) }{ \pi(k_1,\theta^{(k_1)})} \;
\dfrac{\pi_{21} g_{k_2}(u_{k_2}) }{ \pi_{12} g_{k_1}(u_{k_1})}
\; \left| \dfrac{\partial T_{k_1\rightarrow k_2}(\theta^{(k_1)},u_{k_1}) }{ \partial
(\theta^{(k_1)},u_{k_1})} \right|, 1 \right)\;,
$$
involving 
\begin{itemize}
\item[--] the Jacobian of the transform $T_{k_1\rightarrow k_2}$,,
\item[--] the probability $\pi_{ij}$ of
choosing a jump to ${\cal M}_{k_j}$ while in ${\cal M}_{k_i}$, and 
\item[--] $g_i$, the density of $u_i$. 
\end{itemize}
The acceptance probability for the reverse\index{acceptance probability!reverse move}
move is based on the inverse ratio
if the move from ${\mathfrak M}_{k_2}$ to ${\mathfrak M}_{k_1}$ 
also satisfies (\ref{eq:ze_transf}) with $u_{k_2}\sim g_{k_2}(u_{k_2})$.\footnote{For 
a simple proof that the acceptance probability guarantees that
the stationary distribution is $\pi(k,\theta^{(k)})$, see 
Robert and Casella (2004, Section 11.2.2).}

The pseudo-code representation of Green's algorithm is thus as follows:

\bigskip
\centerline{\bf ---Green's Algorithm---}\nopagebreak[4]
\smallskip
\colorbox{LightGrey}{\makebox[0.9\textwidth][l]{\parbox{0.85\textwidth}{
\medskip
{\sf
\noindent At iteration $t$, if $x^{(t)}=(m,\theta^{(m)})$,}
{\sf \begin{enumerate}
\renewcommand{\theenumi}{\arabic{enumi}.}
\item           Select model ${\mathfrak M}_n$ with probability $\pi_{mn}$
\item           Generate $u_{mn}\sim\varphi_{mn}(u)$ 
\item           Set $(\theta^{(n)},v_{nm})=T_{m\rightarrow n}(\theta^{(m)},u_{mn})$
\item           Take $x^{(t+1)} =( n,\theta^{(n)})$ with probability
$$
\min\left(
\dfrac{\pi(n,\theta^{(n)}) }{ \pi(m,\theta^{(m)})}
\;
\dfrac{\pi_{nm} \varphi_{nm}(v_{nm}) }{ \pi_{mn} \varphi_{mn}(u_{mn})}
\; \left| \dfrac{\partial T_{m\rightarrow n}(\theta^{(m)},u_{mn}) }{ \partial
(\theta^{(m)},u_{mn})} \right|, 1 \right)\;,
$$
and take $x^{(t+1)} =x^{(t)}$ otherwise.
\end{enumerate}
}}}}

\bigskip
Even more than previous methods, the implementation of this algorithm requires a
high degree of skillfulness in picking the right proposals and the appropriate scales.
Indeed, it may be argued that the ultimate dependence of the method on the pairwise
completion schemes prevents an extension of its use by a broader audience.
This ``art of reversible jump MCMC" is illustrated on the two following examples,
extracted from Robert and Casella (2004, Section 14.2.3). 

\begin{figure}[hbtp]
\centerline{\includegraphics[width=6cm,height=7cm,angle=270,draft=false]{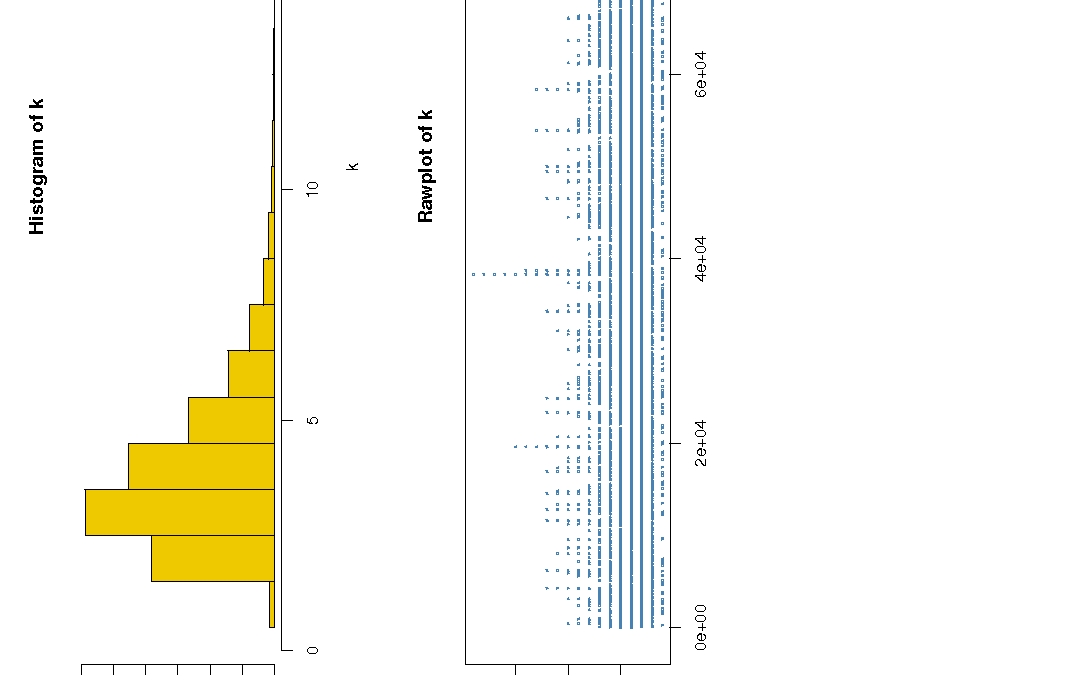}}
\vglue -2truecm
\caption{\label{fig:galtik1}
Histogram and raw plot of $100,000$ $k$'s produced by a reversible jump MCMC 
algorithm for the Galaxy dataset.} 
\end{figure}

\begin{figure}[hbtp]
\begin{center}
\includegraphics[width=10cm,height=12cm,angle=270]{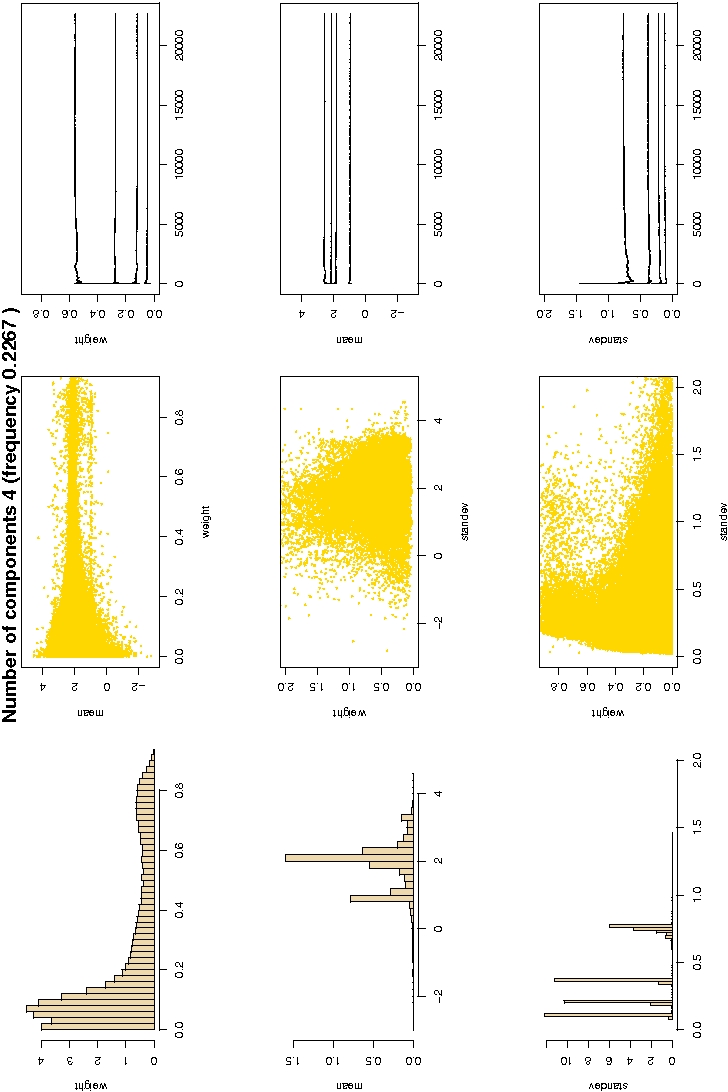}
\end{center}
\caption{\label{fig:galtik2}
Reversible jump MCMC output on the parameters of the model ${\cal M}_3$ for the Galaxy dataset,
obtained by conditioning on $k=3$. {\em The left column gives the histogram of the
weights, means, and variances; the middle column the scatterplot of the pairs
weights-means, means-variances, and variances-weights; the right column plots
the cumulated averages (over iterations) for the weights, means, and variances.}
}
\end{figure}

\begin{figure}[hbtp]
\includegraphics[width=5cm,height=7cm,angle=270]{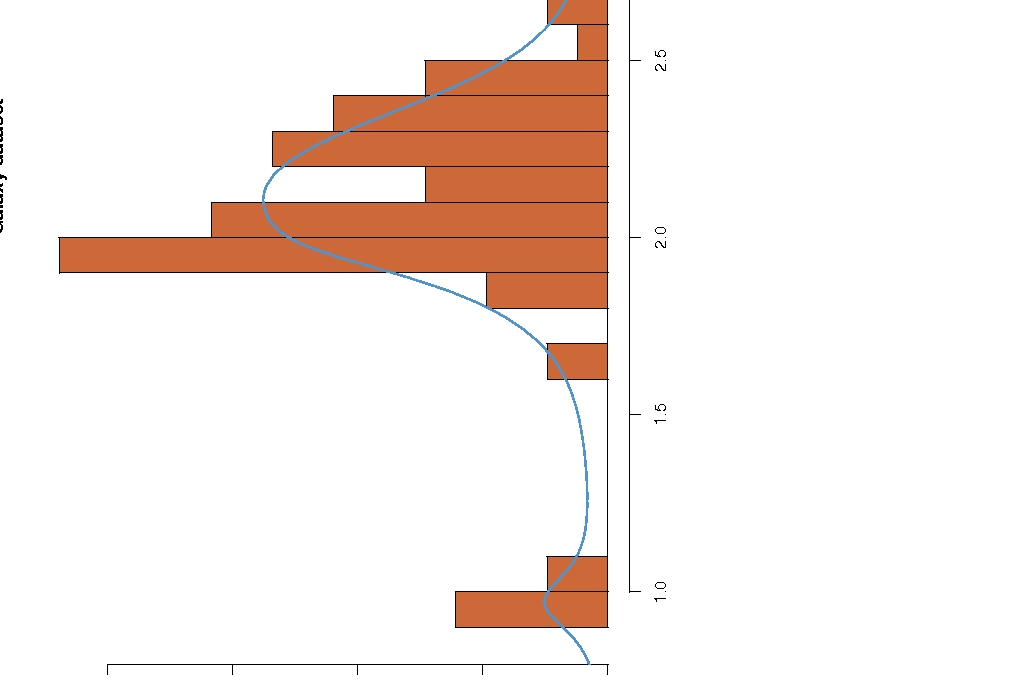}
\vglue -2truecm
\caption{\label{fig:galtik3}
Fit of the dataset by the averaged density, $\mathbb{E}[f(y|\theta)|\mathbf{x}]$ 
}
\end{figure}

\begin{exemple} [Continuation of Example \ref{ex:mix}]\label{ex:RJmix1}
If we consider for model ${\mathfrak M}_k$ the $k$ component normal 
mixture distribution,\index{distribution!mixture}
$$
\sum_{j=1}^k p_{jk} {\cal N}(\mu_{jk},\sigma_{jk}^2)\,,
$$
moves between models involve changing the number of components in the mixture and
thus adding new components or removing older components or yet again changing
several components. As in \cite{Richardson:Green:1997}, we can restrict the moves 
when in model ${\mathfrak M}_k$ to only models
${\mathfrak M}_{k+1}$ and ${\mathfrak M}_{k-1}$. The simplest solution is to use
a birth-and-death process: The {\em birth step} consists in adding a new normal component 
in the mixture generated from the prior and the {\em death step} is the opposite,
removing one of the $k$ components at random. In this case,
the corresponding birth acceptance probability is\index{birth-and-death process}
\begin{align}
&\min\left( \frac{ \pi_{(k+1)k}}{\pi_{k(k+1)} }\,\frac{(k+1)!}{k!}
\frac{\pi_{k+1}(\theta_{k+1})}{\pi_{k}(\theta_{k})\,(k+1)\varphi_{k(k+1)}(u_{k(k+1)})}\,
,1 \right)\nonumber\\
&\quad = \min\left( \frac{ \pi_{(k+1)k}}{\pi_{k(k+1)} }\,
\frac{\varrho(k+1)}{\varrho(k)}\,
\frac{\ell_{k+1}(\theta_{k+1})\,(1-p_{k+1})^{k-1}}{\ell_{k}(\theta_{k})}
,1 \right)\,,\nonumber 
\end{align}
where $\ell_k$ denotes the likelihood of the $k$ component mixture  model
${\mathfrak M}_k$ and $\varrho(k)$ is the prior probability of model
${\mathfrak M}_k$.\footnote{In the birth acceptance probability,
the factorials $k!$ and $(k+1)!$ appear as the
numbers of ways of ordering the $k$ and $k+1$ components of the mixtures.
The ratio cancels with $1/(k+1)$, which is the probability of
selecting a particular component for the death step.}

While this proposal can work well in some setting, as in Richardson and Green (1997)
when the prior is calibrated against the data, it can also be inefficient, that is,
leading to a high rejection rate, if the prior is vague, since the birth proposals are
not tuned properly. A second proposal, central to the solution of 
\cite{Richardson:Green:1997}, is to devise more local jumps between models, called
{\em split} and {\em combine} moves, since a new component is created by splitting
an existing component into two, under some moment preservation conditions, and the
reverse move consists in combining two existing components into one, with symmetric
constraints that ensure reversibility. \citep[See, e.g.,][for details.]{robert:casella:2004}

Figures \ref{fig:galtik1}--\ref{fig:galtik3}
illustrate the implementation of this algorithm for the so-called Galaxy dataset
used by \cite{Richardson:Green:1997} \citep[see also][]{Roeder:1992}, which contains
$82$ observations on the speed of galaxies. On
Figure \ref{fig:galtik1}, the MCMC output on the number of components $k$ is
represented as a histogram on $k$, and the corresponding sequence of $k$'s.
The prior used on $k$ is a uniform distribution on $\{1,\ldots,20\}$: as shown
by the lower plot, most values of $k$ are explored by the reversible jump algorithm,
but the upper bound does not appear to be restrictive since the $k^{(t)}$'s
hardly ever reach this upper limit. 
Figure \ref{fig:galtik2} illustrates the fact that conditioning the output on the 
most likely value of $k$ ($3$ here) is possible. The nine graphs in this Figure show the
joint variation of the three types of parameters, as well as the stability of the
Markov chain over the $1,000,000$ iterations: the cumulated averages are quite 
stable, almost from the start.

The density plotted on top of the histogram in 
Figure \ref{fig:galtik3} is another good illustration of the inferential possibilities offered
by reversible jump algorithms, as a case of {\em model averaging}\index{model!averaging}:
this density is obtained as the average over iterations $t$ of
$$
\sum_{j=1}^{k^{(t)}} p^{(t)}_{jk} {\cal N}(\mu^{(t)}_{jk},(\sigma^{(t)}_{jk})^2)\,,
$$
which approximates the posterior expectation $\mathbb{E}[f(y|\theta)|\mathbf{x}]$,
where $\mathbf{x}$ denotes the data $x_1,\ldots,x_{82}$.\end{exemple}

\begin{exemple}[Continuation of Example \ref{ex:AR}]\label{ex:Jjarjar} \ 
For the $AR(p)$ model of Example \ref{ex:AR}, the best way to include the
stationarity constraints is to use the lag-polynomial representation
$$
\prod_{i=1}^p \left( 1 - \lambda_i B\right) \,X_t = \epsilon_t \,,
\quad \epsilon_t\sim{\mathcal N}(0,\sigma^2)\,,
$$
of model ${\mathfrak M}_p$,\index{AR model}\index{model!AR}
and to constrain the inverse roots, $\lambda_i$, to stay within the unit circle 
if complex and within $[-1,1]$ if real \citep[see, e.g.][Section 4.5.2]{robert:2007}.
The associated uniform priors for the real and complex roots $\lambda_j$ is
$$
\pi_p(\boldsymbol{\lambda}) = \frac{1}{\lfloor p/2 \rfloor+1}\,
\prod_{\lambda_i\in{\mathbb R}} \frac{1}{2}{\mathbb I}_{|\lambda_i|<1}
\,\prod_{\lambda_i\not\in{\mathbb R}} \frac{1}{\pi}{\mathbb I}_{|\lambda_i|<1}\,,
$$
where $\lfloor p/2 \rfloor+1$ is the number of different values of $r_p$.
This factor must be included
within the posterior distribution when using reversible jump since it does not vanish
in the acceptance probability of a move between models ${\mathfrak M}_p$ and ${\mathfrak M}_q$. 
Otherwise, this results in a modification of the prior probability of each model.

Once again, a simple choice is to use a birth-and-death scheme where the birth moves
either create a real or two conjugate complex roots.
As in the birth-and-death proposal for Example \ref{ex:RJmix1}, the acceptance
probability simplifies quite dramatically since it is for instance
$$
\min\left( \frac{\pi_{(p+1)p}}{\pi_{p(p+1)}}\,
\frac{(r_{p}+1)!}{r_{p}!}\,
\frac{\lfloor p/2 \rfloor+1}{\lfloor (p+1)/2 \rfloor+1}
\frac{\ell_{p+1}(\theta_{p+1})}{\ell_{p}(\theta_{p})} ,1 \right)
$$
in the case of a move from ${\mathfrak M}_p$ to ${\mathfrak M}_{p+1}$. (As for the above mixture example,
the factorials are related to the possible choices of the created and the deleted roots.)

\begin{figure}[hbtp]
\centerline{\includegraphics[height=\textwidth,width=9cm,angle=270]{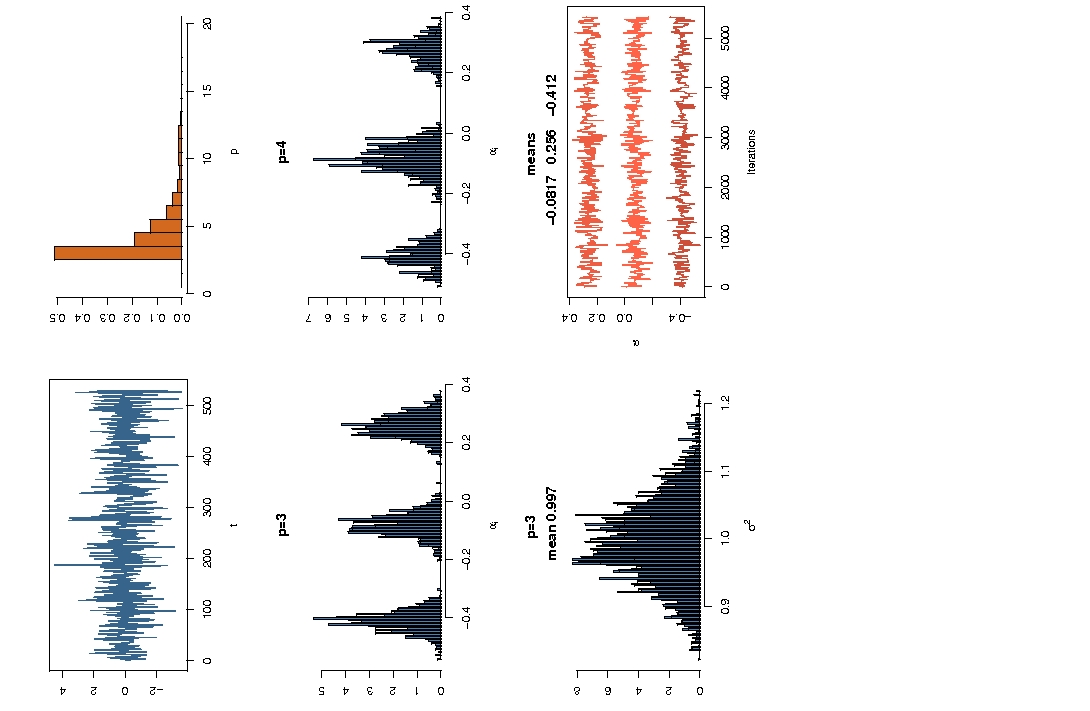}}
\vglue -2.0truecm
\caption[Reversible jump for the $AR(p)$ model]{\label{myfig}
Output of a reversible jump algorithm based on 
an $AR(3)$ simulated dataset of 530 points {\em (upper left)} 
with true parameters $\theta_i$ $(-0.1,0.3,-0.4)$ and $\sigma=1$.
The first histogram is associated with $k$, the following
histograms are associated with the $\theta_i$'s, for
different values of $k$, and of $\sigma^2$. The final graph is a scatterplot
of the complex roots (for iterations where there were complex roots). 
The one before last graph plots the evolution over the iterations of 
$\theta_1,\theta_2,\theta_3$ {\em (Source: Robert 2003)}.
}
\end{figure}

Figure \ref{myfig} presents some views of the corresponding reversible jump MCMC algorithm.
Besides the ability of the algorithm to explore a range of values of $k$, it also shows that
Bayesian inference using these tools is much richer, since it can,
for instance, condition on or average over
the order $k$, mix the parameters of different models and run various tests on these
parameters. A last remark on this graph is that both the order and the value of the
parameters are well estimated, with a
characteristic trimodality on the histograms of the $\theta_i$'s, even when conditioning
on $k$ different from $3$, the value used for the simulation. 
\end{exemple}\index{reversible jump MCMC|)}

In conclusion, while reversible jump MCMC is a generic and powerful method for handling model
comparison when faced with a multitude of models, its sensitivity to the tuning ``parameters"
that determine the pairwise jumps makes us favour the alternative of computing Bayes factors.
When the number of models under comparisons is sufficiently small to allow for the computation
of all marginal likelihoods in parallel, this computational solution is more efficient. It is
indeed quite rare that the structure of the posterior under a model $\mathfrak{M}_1$ provides
information about the structure of the posterior under another model $\mathfrak{M}_2$, unless
both models are quite close. Intrinsically, reversible jump MCMC is a random walk over the collection
of models and, as such, looses in efficiency in its exploration of this collection. (In particular, 
it almost never makes sense to implement reversible jump MCMC when comparing a small number of models.)
\index{Markov chain Monte Carlo (MCMC) algorithm|)}

\section{More Monte Carlo Methods}\label{sec:MoMa}
While MCMC algorithms considerably expanded the range of applications of
Bayesian analysis, they are not, by any means, the end of the story! Further
developments are taking place, either at the fringe of the MCMC realm or far
away from it. We indicate below a few of the directions in Bayesian computational
Statistics, omitting many more that also are of interest...

\subsection{Adaptivity for MCMC algorithms}\label{sub:MoMa:dapt}
Given the range of situations where MCMC applies, it is unrealistic to
hope for a {\em generic} MCMC sampler that would function in every possible
setting. The more generic proposals like random-walk Metropolis--Hastings
algorithms are known to fail in large dimension and disconnected supports,\index{dimension!high}
because they take too long to explore the space of interest \citep{Neal:2003}.
The reason for this impossibility theorem is that, in realistic problems, the
complexity of the distribution to simulation is the very reason why MCMC is
used! So it is difficult to ask for a prior opinion about this distribution,
its support or the parameters of the proposal distribution used in the MCMC
algorithm: intuition is close to void in most of these problems.\index{adaptivity} 

However, the performances of off-the-shelve algorithms like the
random-walk Metropolis--Hastings scheme bring information about 
the distribution of interest and, as such, should be incorporated
in the design of better and more powerful algorithms. The problem 
is that we usually miss the time to train the algorithm on these
previous performances and are looking for the Holy Grail of
automated MCMC procedures! While it is natural to think that the
information brought by the first steps of an MCMC algorithm should
be used in later steps, there is a severe catch: using the whole
past of the ``chain" implies that this is not a Markov chain any
longer. Therefore, usual convergence theorems do not apply and the
validity of the corresponding algorithms is questionable. Further,
it may be that, in practice, such algorithms do degenerate to
point masses because of a too rapid decrease in the variation of
their proposal.\index{Markov chain Monte Carlo (MCMC) algorithm!adaptive}

\begin{figure}[hbtp]
\includegraphics[width=11cm,height=9cm]{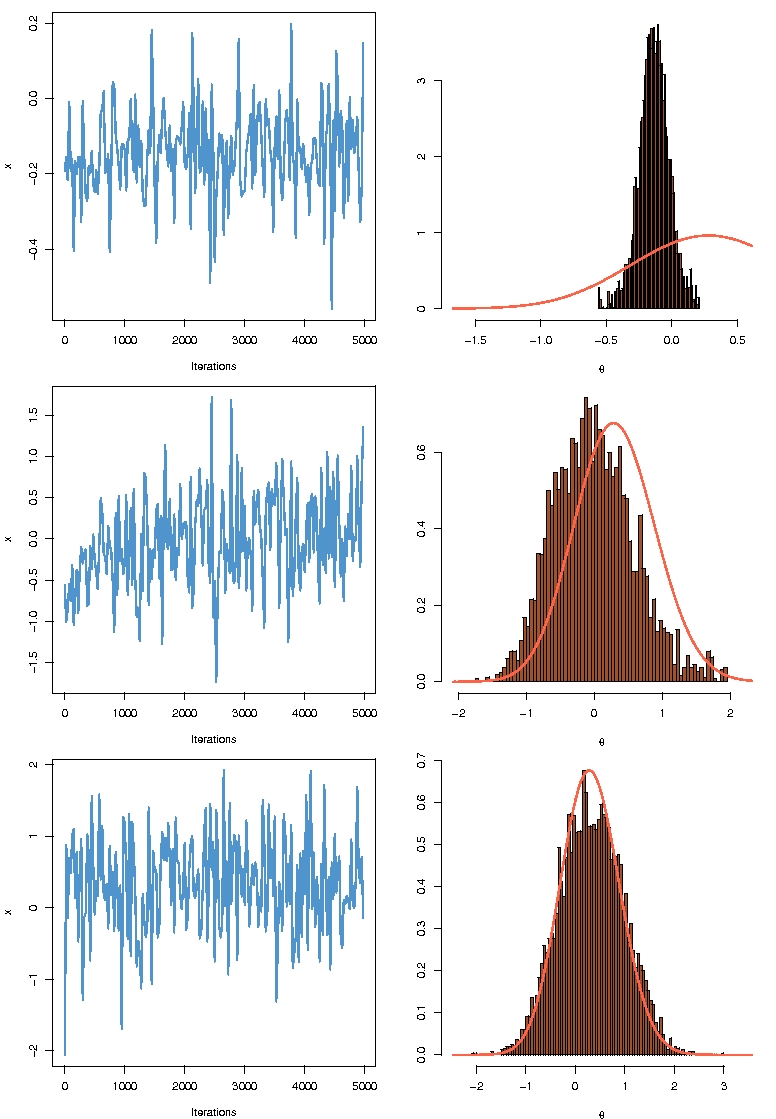}
\caption[Invalid adaptive MCMC]{\label{fig:badap}
Output of the adaptive scheme for the $t$-distribution posterior with a
sample of $10$ $x_j\sim\mathcal{T}_3$ and initial variances of {\em (top)} 0.1, {\em (middle)}
0.5, and {\em (bottom)} 2.5. The left column plots the sequence of $\theta^{(i)}$'s
while the right column compares its histogram against the true posterior distribution
{\em (with a different scale for the upper graph)}.
}
\end{figure}

\begin{figure}[hbtp]
\includegraphics[width=11cm,height=4cm]{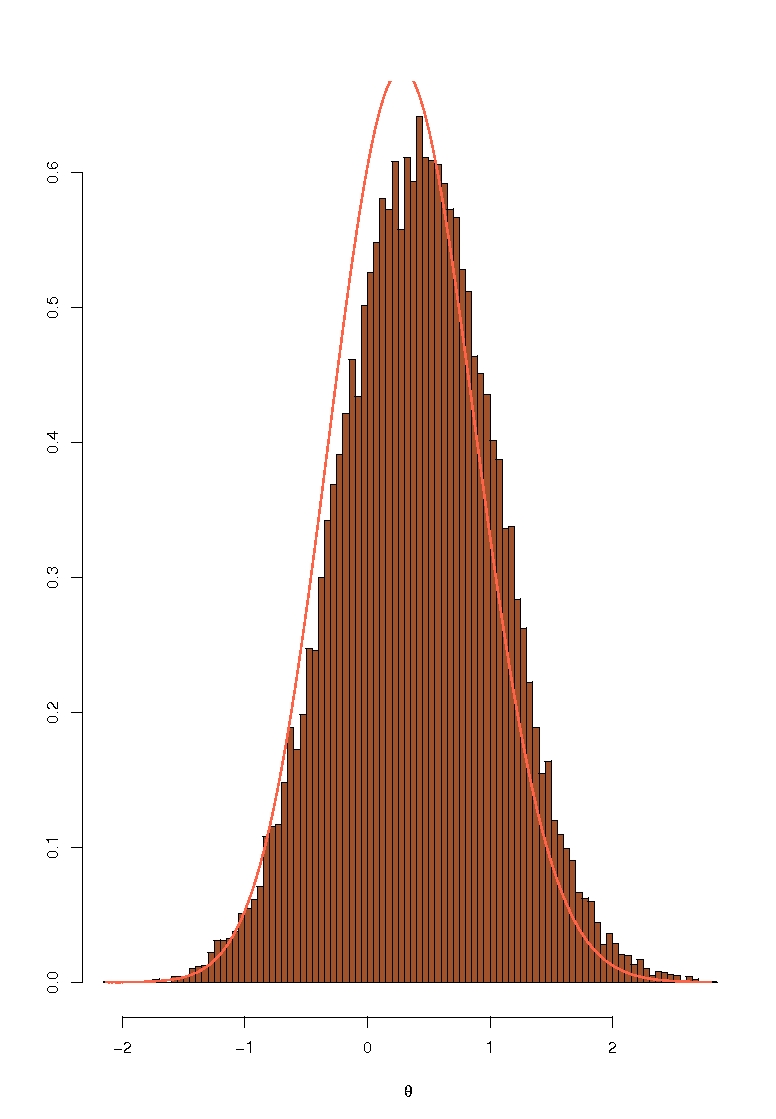}
\caption[Invalid adaptive MCMC with large variance]{\label{fig:badap2}
Comparison of the distribution of an adaptive scheme sample 
of $25,000$ points with initial variance of $2.5$
and of the target distribution.}
\end{figure}

\begin{exemple}[Continuation of Example \ref{ex:toy}]\label{ex:adatoy}
For the $t$-distribution sample, we could fit a normal proposal from the
empirical mean and variance of the previous values of the chain,
$$
\mu_t = \frac{1}{t}\,\sum_{i=1}^t \theta^{(i)}
\quad\hbox{and}\quad
\sigma^2_t = \frac{1}{t}\,\sum_{i=1}^t (\theta^{(i)} - \mu_t)^2\,.
$$
This leads to a Metropolis--Hastings algorithm with acceptance probability
$$
\prod_{j=2}^n \left[ \frac{\nu + (x_j-\theta^{(t)})^2}{\nu + (x_j-\xi)^2}
\right]^{-(\nu+1)/2} \, \frac{\exp -(\mu_t-\theta^{(t)})^2/2\sigma^2_t}{
\exp -(\mu_t-\xi)^2/2\sigma^2_t}\,,
$$
where $\xi$ is the proposed value from $\mathcal{N}(\mu_t,\sigma^2_t)$.
The invalidity of this scheme (because of the dependence on the whole sequence
of $\theta^{(i)}$'s till iteration $t$) is illustrated in Figure \ref{fig:badap}:
when the range of the initial values is too small, the sequence of $\theta^{(i)}$'s
cannot converge to the target distribution and concentrates on too small a support.
But the problem is deeper, because even when the range of the simulated values is
correct, the (long-term) dependence on past values modifies the distribution of the
sequence. Figure \ref{fig:badap2} shows that, for an initial variance of $2.5$, there
is a bias in the histogram, even after $25,000$ iterations and stabilisation of the
empirical mean and variance.\index{adaptivity!invalid}
\end{exemple}

\begin{figure}[hbtp]
\includegraphics[width=11cm,height=6cm]{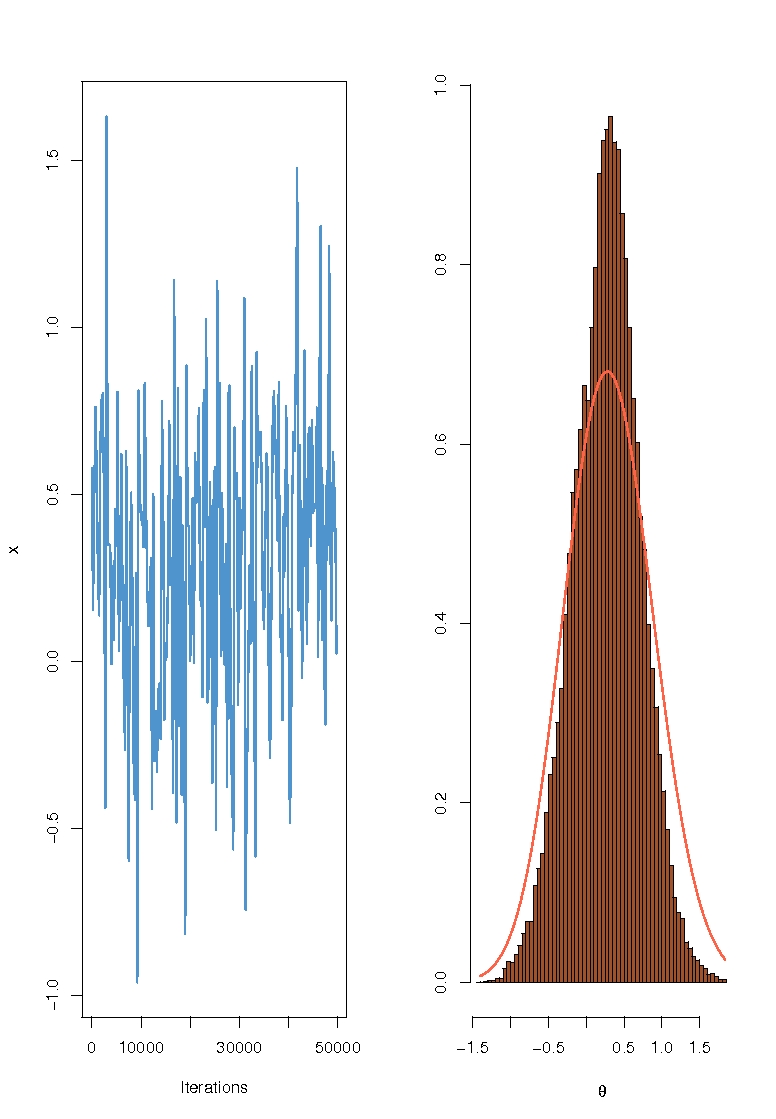}
\caption[Nonparametric adaptive MCMC]{\label{fig:nopada}
Sample produced by $50,000$ iterations of a nonparametric adaptive MCMC scheme and
comparison of its distribution with the target distribution.}
\end{figure}

Even though the Markov chain is converging {\em in distribution} to the
target distribution (when using a proper, i.e.~time-homogeneous updating
scheme), using past simulations to create a non-parametric approximation
to the target distribution does not work either. Figure \ref{fig:nopada}
shows for instance the output of an adaptive scheme in the setting of 
Example \ref{ex:adatoy} when the proposal distribution is the Gaussian
kernel based on earlier simulations. A very large number of iterations
is not sufficient to reach an acceptable approximation of the target distribution.

The overall message is thus that, without further guidance,
one should not {\em constantly} adapt the proposal
distribution on the past performances of the simulated chain. Either the adaptation
must cease after a period of {\em burnin} (not to be taken into account for the
computations of expectations and quantities related to the target distribution).
Else, the adaptive scheme must be theoretically assess on its own right. This later path is
not easy and only a few examples can be found (so far) in the literature. See, e.g.,
\cite{Gilks:Roberts:Sahu:1998} who use regeneration to create block independence and
preserve Markovianity on the paths rather than on the values, \cite{Haario:Sacksman:Tamminen:1999,
Haario:Sacksman:Tamminen:2001} who derive a proper adaptation scheme in the spirit of 
Example \ref{ex:adatoy} by using a ridge-like correction to the empirical
variance, and \cite{Andrieu:Robert:2001} who propose a more general
framework of valid adaptivity based on stochastic optimisation and the Robbin-Monro
algorithm.\index{adaptivity} A more generic perspective is found in
\cite{roberts:rosenthal:2006}, who tune the random walk scale in each direction of the parameter space toward an
optimal acceptance rate of $0.44$, a rate suggested in \cite{gelman:gilks:roberts:1996}.\index{acceptance rate!optimal}
\cite{roberts:rosenthal:2006} provide validated constraints on the tuning scheme to the extent of 
offering an {\sf R} package called \verb+amcmc+ and described in \cite{rosenthal:2007}.
More precisely, for each component of the simulated parameter,
a factor $\delta_i$ corresponding to the logarithm of the random
walk standard deviation is updated every $50$ iterations by adding or subtracting a factor
$\epsilon_t$ depending on whether or not the average acceptance rate on that batch of
$50$ iterations and for this component was above or below $0.44$. If $\epsilon_t$ decreases to
zero as $\min(.01,1/\sqrt{t})$, the conditions for convergence are satisfied.\index{amcmc@\verb+amcmc+}
(See Robert and Casella (2009, Section 8.5.2, for more details.)

\subsection{Population Monte Carlo}\label{sub:MoMa:pOp}
To reach acceptable adaptive algorithms, while avoiding an extended study of
their theoretical properties, a better alternative is to leave the setup
of Markov chain simulations and to consider instead {\em sequential} or {\em population} Monte 
Carlo methods \citep{iba00,Cappe:Guillin:Marin:Robert:2003}\index{Monte Carlo techniques!sequential} 
that have much more in common with importance sampling than with MCMC. They
are inspired from {\em particle systems} that were introduced to handle 
rapidly\index{particle systems}\index{Monte Carlo techniques!population} 
changing target distributions like those found in signal processing and imaging 
\citep{Gordon:Salmon:Smith:1993,Shephard:Pitt:1997,Doucet:deFreitas:Gordon:2001,delmoral:doucet:jasra:2006} 
but primarily handle fixed but complex target distributions by building a sequence of 
increasingly better proposal distributions.\footnote{The ``sequential" denomination
in the sequential Monte Carlo methods\index{population Monte Carlo (PMC) techniques}
thus refers to the algorithmic part, not to the statistical part.} Each
iteration of the population Monte Carlo (PMC) algorithm thus produces a 
sample approximately simulated from the target distribution but the iterative 
structure allows for adaptivity toward the target distribution.\index{adaptivity} Since
the validation is based on importance sampling principles, dependence on
the past samples can be arbitrary {\em and} the approximation to the target
is valid (unbiased) at {\em each iteration} and does not require convergence 
times nor stopping rules.\index{stopping rule} 

If $t$ indexes the iteration and $i$ the sample point, consider proposal distributions $q_{it}$
that simulate the $x_i^{(t)}$'s and associate to each \smash{$x_i^{(t)}$} an importance weight
$$
\varrho_i^{(t)} = \pi(x_i^{(t)}) \big/ q_{it}(x_i^{(t)})\,, \qquad i=1,\ldots,n\,.
$$
(The proposal distribution thus depends both on the iteration and on the particle index.)
Approximations of the form
$$
\mathfrak{I}_t = \frac{1}{n}\,\sum_{i=1}^n \varrho_i^{(t)} \,h(x_i^{(t)})
$$
are then unbiased estimators of $\mathbb{E}^\pi[h(X)]$, even when
the importance distribution $q_{it}$ depends on the entire past of the experiment. Indeed, 
if $\zeta$ denotes the vector of past random variates that contribute to $q_{it}$, 
and $g(\zeta)$ its {\em arbitrary} distribution, we have
$$
\int\int \frac{\pi(x)}{q_{it}(x|\zeta)} \, h(x) q_{it}(x) \text{d}x\,g(\zeta) \text{d}\zeta 
= \int\int h(x) \pi(x) \text{d}x\,g(\zeta) \text{d}\zeta 
= \mathbb{E}^\pi[h(X)]\,.
$$
Furthermore, assuming that the variances
$${\mathrm{var}}\left(\varrho_i^{(t)} h(x_i^{(t)})\right)$$ exist for every $1\le i\le  n$, we have
$$
{\mathrm{var}}\left( \mathfrak{I}_t \right) =  \frac{1}{n^2}\ \sum_{i=1}^n 
{\mathrm{var}}\left(\varrho_i^{(t)} h(x_i^{(t)})\right) \,,
$$
due to the cancelling effect of the weights $\varrho_i^{(t)}$.

Since, usually, the density $\pi$ is unscaled, we use instead
$$
\varrho_i^{(t)} \propto \frac{\pi(x_i^{(t)})}{q_{it}(x_i^{(t)})}\,,
\qquad i=1,\ldots,n\,,
$$
scaled so that the $\varrho_i^{(t)}$'s sum up to $1$. In this case, the
unbiasedness is lost, although
it approximately holds. In fact, the estimation of the normalising constant of
$\pi$ improves with each iteration $t$, since the overall average
$$
\varpi_t = \frac{1}{tn}\,\sum_{\tau=1}^t\,\sum_{i=1}^n\,
\frac{\pi(x_i^{(\tau)})}{q_{i\tau}(x_i^{(\tau)})}
$$
is convergent. Therefore, as $t$ increases, $\varpi_t$ contributes less and less 
to the variability of $\mathfrak{I}_t$.

However, \cite{douc:guillin:marin:robert:2005} have shown that this use of the importance
weights $\varrho_i^{(t)}$ in \cite{Cappe:Guillin:Marin:Robert:2003} was not adaptive enough and they proposed
a Rao--Blackwellised substitute\index{Rao--Blackwellisation}\footnote{The generic Rao--Blackwellised improvement
was introduced in the original MCMC paper of \cite{Gelfand:Smith:1990} and studied by \cite{liu:wong:kong:1994}
and \cite{casella:robert:1996}. More recent developments are proposed in \cite{cornuet:marin:mira:robert:2009},
in connection with adaptive algorithms like PMC.}
$$
\varrho_i^{(t)} \propto \frac{\pi(x_i^{(t)})}{\sum_j q_{jt}(x_i^{(t)})}\,,
\qquad i=1,\ldots,n\,,
$$
that guaranteed an (asymptotic in $n$) improvement of the proposal at each iteration $t$, under specific
forms of the $q_{it}$'s \citep{douc:guillin:marin:robert:2005,douc:guillin:marin:robert:2007b,cappe:douc:guillin:marin:robert:2007}

Since the above establishes that an simulation scheme based on
sample dependent proposals is fundamentally a specific kind of importance
sampling, the following algorithm is validated by the same
principles as regular importance sampling:

\medskip
\centerline{\bf ---Population Monte Carlo Algorithm---}\nopagebreak[4]
\smallskip
\colorbox{LightGrey}{\makebox[0.9\textwidth][l]{\parbox{0.85\textwidth}{ 
\medskip
{\sf
\noindent For $t=1,\ldots,T$
\begin{enumerate}
\renewcommand{\theenumi}{\arabic{enumi}.}
\item For $i=1,\ldots,n$,
\begin{enumerate}
\renewcommand{\theenumii}{\roman{enumii}}
\item\label{zistep} Select the generating distribution $q_{it}(\cdot)$
\item Generate $x_i^{(t)}\sim q_{it}(x)$
\item Compute $\varrho_i^{(t)} = \pi(x_i^{(t)})/\sum_j q_{jt}(x_i^{(t)})$
\end{enumerate}
\item Normalise the $\varrho_i^{(t)}$'s to sum up to $1$
\item Resample $n$ values from the $x_i^{(t)}$'s with replacement,
using the weights $\varrho_i^{(t)}$, to create the sample $(x_1^{(t)},\ldots,x_n^{(t)})$
\end{enumerate}
}
}}}

\bigskip
Step (i) is singled out because it is the central property
of the PMC algorithm, namely that adaptivity can be extended to the 
individual level and that the $q_{it}$'s can be picked based on the performances
of the previous $q_{i(t-1)}$'s or even on all the previously simulated
samples, if storage allows. For instance, the $q_{it}$'s can
include large tails proposals as in the {\em defensive sampling} strategy of
\cite{Hesterberg:1998}, to ensure finite variance. Similarly, Warnes' (2001)
non-parametric Gaussian kernel approximation can be used as a proposal.\footnote{Using
a Gaussian non-parametric kernel estimator amounts to (a) sampling from the \smash{$x_i^{(t)}$}'s
with equal weights and (b) using a normal random walk move from the selected 
\smash{$x_i^{(t)}$}, with standard deviation equal to the bandwidth of the kernel.}
(See also in \citealp{Stavropoulos:Titterington:2001} the {\em smooth bootstrap} 
technique as an earlier example of PMC algorithm.)\index{defensive sampling}\index{smooth
bootstrap}\index{bootstrap!smooth}\index{sampling!defensive}

A major difference between the PMC algorithm and earlier proposals in the 
particle system literature is that past dependent moves as those of \cite{Gilks:Berzuini:2001}
and \cite{delmoral:doucet:jasra:2006}
remain within the MCMC framework, with Markov transition kernels with stationary distribution
equal to $\pi$. 

\begin{exemple}[Continuation of Example \ref{ex:mix2}]\label{ex:PMix}
We consider here the implementation of the PMC algorithm in the case of the
the normal mixture (\ref{eq:gimi9}). As in Example \ref{ex:mix3}, 
a PMC sampler can be efficiently implemented {\em without} the (Gibbs) augmentation 
step, using normal random walk proposals based on the previous sample of
$\boldsymbol{\mu}=(\mu_1,\mu_2)$'s. Moreover, the difficulty inherent
to random walks, namely the selection of a ``proper" scale,\index{proposal!multiscale}
can be bypassed because of the adaptivity of the PMC algorithm. Indeed, the
proposals can be associated with a range of variances $v_k$ $(1\le k\le K)$ 
ranging from, e.g., $10^3$ down to $10^{-3}$. At each step of the algorithm, 
the new variances can be selected proportionally to the performances of the scales 
$v_k$ on the previous iterations. For instance, a scale can be chosen proportionally 
to its {\em non-degeneracy rate} in the previous iteration, that is, the percentage of
points generated with the scale $v_k$ that survived after resampling.\footnote{When
the survival rate of a proposal distribution is null, in order to avoid the complete 
removal of a given scale $v_k$, the corresponding number $r_k$ of proposals with that scale
is set to a positive value, like $1$\%~of the sample\index{survival rate!variance} 
size.} The weights are then of the form\index{importance sampling!degeneracy of}\index{survival rate!null}
$$
\varrho_i\propto\frac{f\left(\mathbf{x}\left|\left(\mu_1\right)_i^{(t)},\left(\mu_2\right)_i^{(t)}\right.\right)
\pi\left(\left(\mu_1\right)_i^{(t)},\left(\mu_2\right)_i^{(t)}\right)}
{\sum_{j=1}^n \varphi\left(\left(\mu_1\right)_i^{(t)}\left|\left(\mu_1\right)_j^{(t-1)},v_k\right.\right)
\varphi\left(\left(\mu_2\right)_i^{(t)}\left|\left(\mu_2\right)_j^{(t-1)},v_k\right.\right)\,,}
$$
where $\varphi(q|s,v)$ is the density of the normal distribution with
mean $s$ and variance $v$ at the point $q$.

Compared with an MCMC algorithm in the same setting (see Examples \ref{ex:mix2}
and \ref{ex:mix3}), the main feature of this algorithm is its ability to deal
with multiscale proposals in an unsupervised manner. The upper row of Figure 
\ref{fig:pmcmix2} produces the frequencies of the five variances $v_k$ used 
in the proposals along iterations: The two largest variances $v_k$ most often 
have a zero survival rate, but sometimes experience bursts of survival.
In fact, too large a variance mostly produces points that are irrelevant for 
the posterior distribution, but once in a while a point $\theta_j^{(t)}$ gets
close to one of the modes of the posterior. When this occurs, the corresponding
$\varrho_j$ is large and $\theta_j^{(t)}$ is thus heavily resampled.
The upper right graph shows that the other proposals are rather evenly sampled along
iterations. The influence of the variation in the proposals on the estimation
of the means $\mu_1$ and $\mu_2$ can be seen on the middle and lower panels of
Figure \ref{fig:pmcmix2}. First, the cumulative averages quickly stabilise
over iterations, by virtue of the general importance sampling proposal.
Second, the corresponding variances take longer to stabilise
but this is to be expected, given the regular reappearance of
subsamples with large variances.\index{PMC vs.~MCMC}

\begin{figure}[hbtp]
\begin{center}
\includegraphics[height=7cm,width=12cm]{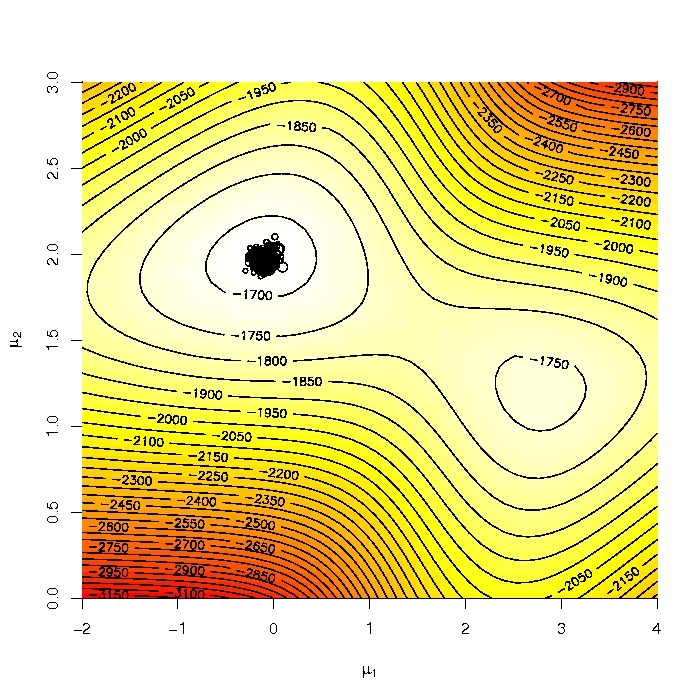}
\caption{Representation of the log-posterior distribution 
with the PMC weighted sample after $30$ iterations (the weights are proportional to
the circles at each point) {\em (Source: \citealp{Cappe:Guillin:Marin:Robert:2003})}.}
\label{fig:pmc2mix}
\end{center}
\end{figure}

\begin{figure}[hbtp]
\begin{center}
\includegraphics[height=6cm,width=11cm]{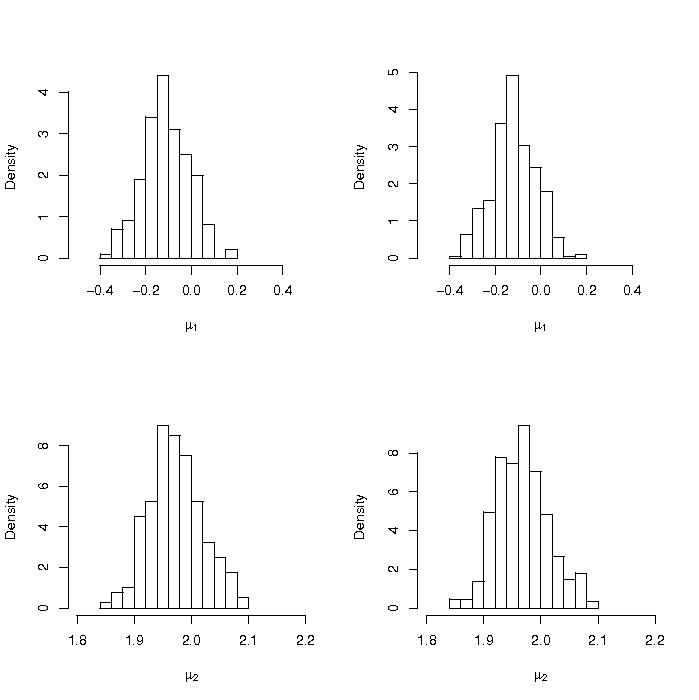}
\caption{Histograms of the PMC sample: sample at iteration 5 {\em (left)}
before resampling and {\em (right)} after resampling.}
\label{fig:hpmc2mix}
\end{center}
\end{figure}

\begin{figure}[hbtp]
\includegraphics[width=11cm,height=10cm,draft=false]{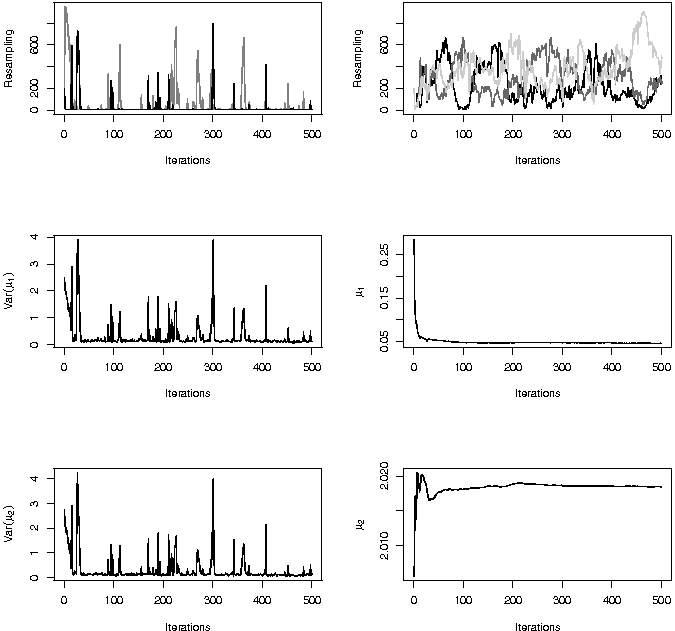}
\caption[Output from mixture PMC algorithm]{Performances of the mixture PMC algorithm
for $1000$ observations from a $0.2 {\cal N}(0,1)+0.8{\cal N}(2,1)$ distribution,
with $\theta=1$ $\lambda=0.1$, $v_k=5,2,.1,.05,.01$, and a population of $1050$ particles:
{\em (upper left)} Number of resampled points for the variances $v_1=5$ (darker) and $v_2=2$;
{\em (upper right)} Number of resampled points for the other variances, $v_3=0.1$ is the darkest 
one; {\em (middle left)} Variance of the simulated $\mu_1$'s along iterations;
{\em (middle right)} Cumulated average of the simulated $\mu_1$'s over iterations;
{\em (lower left)} Variance of the simulated $\mu_2$'s along iterations; 
{\em (lower right)} Cumulated average of the simulated $\mu_2$'s over iterations
{\em (Source: \citealp{Cappe:Guillin:Marin:Robert:2003})}.
\label{fig:pmcmix2}}
\end{figure}

In comparison with Figures \ref{fig:wronmix} and \ref{fig:rwmix},
Figure \ref{fig:pmc2mix} shows that the sample produced by the PMC algorithm 
is quite in agreement with the modal zone of the posterior distribution. 
The second mode, which is much lower, is not preserved in the sample after 
the first iteration.  Figure \ref{fig:hpmc2mix} also shows that the weights 
are quite similar, with no overwhelming weight in the sample.
\end{exemple}

The generality in the choice of the proposal distributions $q_{it}$ is 
obviously due to the abandonment of the MCMC framework.\index{proposal!adaptive}
The difference with an MCMC framework is not simply a theoretical advantage: 
as seen in Section \ref{sub:MoMa:dapt}, proposals based on the whole past of
the chain do not often work. Even algorithms validated by MCMC steps
may have difficulties: in one example of \cite{Cappe:Guillin:Marin:Robert:2003},
a Metropolis--Hastings scheme does not work well, while a PMC algorithm  
based on the same proposal produces correct answers.
Population Monte Carlo algorithms offer a theoretically valid solution to the
adaptivity issue, with practical gains as well, as exemplified in \cite{wraith-2009-80}.
The connection with nonparametric Bayes estimation has not yet been sufficiently pursued
but the convergence of kernel-like proposals demonstrated by \cite{cappe:douc:guillin:marin:robert:2007}
shows this is a promising direction.

\subsection{Approximate Bayesian computation}\label{sec:abc}
One particular application of the Accept--Reject algorithm that has found a niche of its own
in population genetics is called approximate Bayesian computation (ABC)\index{ABC}\index{approximate 
Bayesian computation (ABC)}, following the denomination proposed by 
\cite{pritchard:seielstad:perez:feldman:1999}. It is in
fact an appealing substitute for ``exact" (meaning convergent) Bayesian algorithms in settings where
the likelihood function $f(x|\theta)$ is not available, even up to a normalising constant, but where
it can easily be simulated. The range of applications is thus much wider than population genetics and
covers for instance graphical models and inverse problems.\index{likelihood!intractable} 

Assuming, thus, that a posterior distribution $\pi(\theta|x_0)\propto \pi(\theta)f(x_0|\theta)$ is to be
simulated, a rudimentary accept-reject algorithm generates values from the prior and from the likelihood until the
simulated observation is equal to the original observation $x_0$:\\
{\sf Repeat}\\
\hglue 1.5truecm {\sf Generate $\theta\sim\pi(\theta)$ and $X\sim f(x|\theta)$}\\
{\sf until $X=x_0$}\\
Since the conditional probability of acceptance is $f(x_0|\theta)$, 
the distribution of the accepted $\theta$ is truly $\pi(\theta|x_0)$, even when $f(x|\theta)$
cannot be computed.

The above algorithm is valid, but it is unfortunately restricted to settings where (a) $\pi(\theta)$ is a proper
prior and (b) $\text{Pr}_\theta(X=x_0)$ has a positive probability of occurrence. Even in population genetics 
where the outcome $X$ is a discrete random variable, the size of the state-space is often such that this 
algorithm cannot be implemented. The proposal
of \cite{pritchard:seielstad:perez:feldman:1999} is to replace the {\em exact} acceptance
condition $X=x_0$ in the above with an {\em approximate} condition $d(X,x_0)<\epsilon$, where $d$ is
a distance and $\epsilon$ is a tolerance level. The corresponding ABC algorithm is thus:

\bigskip
\centerline{\bf ---Approximate Bayesian computation Algorithm---}\nopagebreak[4]
\smallskip
\colorbox{LightGrey}{\makebox[0.9\textwidth][l]{\parbox{0.85\textwidth}{
\medskip
{\sf
For $i=1,\ldots,n$,
\begin{enumerate}
\renewcommand{\theenumii}{\roman{enumii}}
\item Generate $\theta_i\sim \pi(\theta)$
\item Generate $x_i\sim f(x|\theta_i)$ and compute $d(x_0,x_i)$
\end{enumerate}
Accept the $\theta_i$'s such that $d(x_i,x)<\epsilon$.
}
}}}

\bigskip
The output of the ABC algorithm is distributed from the distribution with density proportional to
        $\pi(\theta)\,\text{Pr}_\theta\{d(X,x_0)<\epsilon\}$,
where $\text{Pr}_\theta$ represents the sampling distribution of $X$, indexed by $\theta$. This density
is somehow mistakenly denoted by $\pi\{\theta\mid d(x,x_0)<\epsilon\}$, where the conditioning corresponds to
the marginal distribution of $d(x,x_0)$ given $x_0$. While unavoidable, this inherent approximation step makes
the ABC method difficult to assess and to compare with regular Monte Carlo approaches, even though 
recent works replace it within a non-parametric framework that aims at approximating the conditional
density function $f(x|\theta)$ and hence envision $\epsilon$ as a potential bandwidth
\citep{beaumont:zhang:balding:2002,blum:francois:2010}. 

Improvements on the original scheme have been achieved either by modifying the proposal distribution of the parameter 
$\theta$, in order to increase the density of $x$'s within the vicinity of $y$ \citep{marjoram:etal:2003,bortot:coles:sisson:2007},
or, as stated above, by envisioning the problem as a conditional density estimation issue and 
by developing techniques to allow for larger $\epsilon$ \citep{beaumont:zhang:balding:2002}. For instance, 
\cite{marjoram:etal:2003} defined a Markov chain Monte Carlo (MCMC) 
version of the ABC algorithm that enjoys the same validity as the original, namely that,
if a Markov chain $(\theta^{(t)})$ is created via the transition function
\[
\theta^{(t+1)}= \begin{cases}
        \theta^\prime\sim K(\theta^\prime\mid\theta^{(t)}) &\text{if }x\sim f(x\mid\theta^\prime)\text{ is such that }x=x_0\\
                &\text{ and }u\sim\mathcal{U}(0,1) \le \dfrac{\pi(\theta^\prime)
                K(\theta^{(t)}\mid\theta^\prime)}{\pi(\theta^{(t)}) K(\theta^\prime\mid\theta^{(t)})}\,,\\
        \theta^{(t)} &\text{otherwise,}
                \end{cases}
\]
its stationary distribution is the posterior $\pi(\theta\mid x_0)$. Again, in most settings, exact equality is
not achievable and the above constraint $x=x_0$ is replaced with the approximation $d(x,x_0)<\epsilon$. In
\cite{beaumont:cornuet:marin:robert:2009}, a population Monte Carlo modification of ABC is introduced, resulting into the
algorithm:

\bigskip
\centerline{\bf ---PMC-ABC Algorithm---}\nopagebreak[4]
\smallskip
\colorbox{LightGrey}{\makebox[0.9\textwidth][l]{\parbox{0.85\textwidth}{
\medskip
{\sf
\noindent Given a decreasing sequence of tolerance thresholds $\epsilon_1\ge\ldots\ge\epsilon_T$,
\begin{itemize}
\item[1.] At iteration $t=1$,
\begin{itemize}
  \item[] For $i=1,...,N$
    \begin{itemize}
    \item[ ]Simulate $\theta_i^{(1)}\sim \pi(\theta)$ and $x\sim f(x\mid \theta_i^{(1)})$ until $\varrho(x,y)<\epsilon_1$
    \item[ ]Set $\omega^{(1)}_i=1/N$
    \end{itemize}
\end{itemize}
Take $\tau^2_1$ as twice the empirical variance of the $\theta_i^{(1)}$'s
\item[2.] At iteration $2\le t\le T$,
\begin{itemize}
  \item[] For $i=1,...,N$, repeat
    \begin{itemize}
    \item[] Pick $\theta_i^\star$ from the $\theta_j^{(t-1)}$'s with probabilities $\omega_j^{(t-1)}$\\
      generate $\theta_i^{(t)}\mid \theta_i^\star \sim \mathcal{N}(\theta_i^\star,\tau_{t}^2)$
      and $x\sim f(x\mid \theta_i^{(t)})$
    \end{itemize}
    until $\varrho(x,y)<\epsilon_t$\\
    Set $\omega^{(t)}_i\propto \pi(\theta^{(t)}_i)/\sum_{j=1}^N \omega^{(t-1)}_j
\varphi\left\{ \tau_t^{-1}\left(\theta^{(t)}_i-\theta^{(t-1)}_j\right)\right\}$
\end{itemize}
Take $\tau_{t+1}^2$ as twice the weighted empirical variance of the $\theta_i^{(t)}$'s
\end{itemize}
}
}}}

\bigskip
From a practical viewpoint, the number of iterations $T$ can be controlled via the modifications
in the parameters of $K_t$, a stopping rule being that the iterations should stop when those parameters
have settled, while the more fundamental issue of selecting a sequence of $\epsilon_t$'s towards a
proper approximation of the true posterior can rely on the stabilisation of the estimators of some
quantities of interest associated with this posterior. But there is generally no fixed-cost solution
that let $\epsilon_t$ decrease to zero with $t$. Note also that the SMC algorithm of \cite{delmoral:doucet:jasra:2006} 
has been recently extended to this ABC setup, making the resulting algorithm a direct competitor of the above PMC-ABC
algorithm.

\section{Conclusion}\label{sec:konk}
This short overview of the problems and solutions considered for
Bayesian Statistics is nothing but an introduction to the game:
there are much more complex problems than those illustrated above
and much more advanced techniques than those presented in these
pages. The reader is then encouraged to enter the literature on
the topic, maybe with other introductory surveys like
\cite{Cappe:Robert:2000} and \cite{Andrieu:Doucet:Robert:2004},
but mostly through books like \cite{Chen:Shao:Ibrahim:2000},
\cite{Doucet:deFreitas:Gordon:2001}, \cite{Liu:2001}, 
\cite{robert:casella:2004}, \cite{albert:2007}, \cite{marin:robert:2007},
and \cite{robert:casella:2009}. 

We have not mentioned so far entries to Bayesian\index{Bayesian!software}
softwares like {\sf winBUGS}, developed by the MRC Unit in Cambridge
\citep{Bugs1.1,Bugs1.2}, {\sf Ox} \citep{ox}, {\sf BATS}\index{winBUGS@{\sf winBUGS}} 
\citep{Bats}, {\sf BACC} \citep{geweke99} and the {\sf Minitab} 
package of \cite{albertab}, which all cover some aspects of Bayesian
computing. Obviously, these packages require some
expertise from the user and are thus more difficult of use than the
classical open source or commercial softwares like {\sf R},
{\sf Splus}, {\sf Statgraphics}, {\sf StatXact},
{\sf SPSS} or {\sf SAS}. In other words, they are not {\em black
boxes} that could be used by laymen with no statistical background. But
this entrance fee to the use of Bayesian softwares is inevitable, given
the versatile nature of Bayesian analysis: since it offers much more
variability than standard inferential procedures, through the choice of
prior distributions and loss functions for instance, it also requires
more input from the user! And, once these preliminary steps have been
overcome, the programming involved in a software like {\sf winBUGS} is
rather limited and certainly not harder than writing a code in {\sf R}
or {\sf Matlab}.\index{R@{\sf R}}\footnote{An {\sf R} package called 
\verb+mcsm+ 
has been developed in association with \cite{robert:casella:2009} for
training about Monte Carlo methods.}\index{mcsm@{\verb+mcsm+}}

As stressed in this Chapter, computational issues are central to the
design and implementation of Bayesian analysis. The new era opened
by the MCMC methodology has brought much more freedom in the use of
Bayesian methods, as reflected by the increase of Bayesian studies
in applied Statistics. As usually the case, a strong increase in the
use of a methodology also sees a corresponding increase in its misuse!
Inconsistent data-dependent priors and improper posteriors are sometimes
appearing in studies and, more generally, the assessment of prior modelling
(or even of MCMC convergence) are rarely conducted with sufficient care.
This is somehow a price to pay for the wider range of Bayesian studies,
while the improvement of corresponding software should bring more guidelines
and warnings about these misuses of Bayesian analysis.

\section*{Acknowledgements}
This work had been partly supported by the Agence Nationale de la Recherche (ANR, 212,
rue de Bercy 75012 Paris) through the 2009-2012 projects {\sf Big'MC} and {EMILE}.

\def\enotesize{\normalsize}
\theendnotes
\addcontentsline{toc}{chapter}{\protect\numberline{}Bibliography}
\backmatter
\bibliographystyle{plainnat}

\printindex
\end{document}